\documentclass[12pt,letterpaper]{article}

\usepackage{amsmath, amsthm,amssymb}
\usepackage{ctable}
\usepackage{adjustbox}
\usepackage{hyphenat}
\usepackage{fullpage}
\usepackage[english]{babel}
\usepackage{amsfonts}
\usepackage{color}
\usepackage{setspace}
\usepackage{lscape}
\usepackage{indentfirst}
\usepackage[normalem]{ulem}
\usepackage{booktabs}
\usepackage{natbib}
\usepackage{float}
\usepackage{latexsym}
\usepackage[hang,flushmargin]{footmisc}
\usepackage{lscape}

\usepackage[breaklinks]{hyperref}
\usepackage[all]{hypcap}

\definecolor{citescol}{RGB}{73,0,165}
\definecolor{urlscol}{RGB}{0,107,124}
\definecolor{linkscol}{RGB}{187,24,0}
\hypersetup{colorlinks=true,linkcolor=linkscol,citecolor=citescol,urlcolor=urlscol}

\newcolumntype{R}[1]{>{\raggedleft\arraybackslash}p{#1}}

\usepackage{newfloat}
\DeclareFloatingEnvironment[name={Supplementary Figure}]{suppfigure}
\DeclareFloatingEnvironment[name={Supplementary Table}]{supptable}

\newcommand{\turnover}{\nu}

\newcommand{\fosp}{s}

\definecolor{darkgreen}{rgb}{0, 0.5, 0}

\raggedright
\renewcommand{\section}[1]{%
\bigskip
\begin{center}
\begin{Large}
\normalfont\scshape #1
\medskip
\end{Large}
\end{center}}

\renewcommand{\subsection}[1]{%
\bigskip
\begin{center}
\begin{large}
\normalfont\itshape #1
\end{large}
\end{center}}

\renewcommand{\subsubsection}[1]{%
\vspace{2ex}
\noindent
\textit{#1.}---}

\renewcommand{\tableofcontents}{}

\bibpunct{(}{)}{;}{a}{}{,}  

\begin{document}
\begin{flushright}
Version dated: \today
\end{flushright}
\bigskip
\noindent RH:  Bayesian total-evidence dating

\bigskip
\medskip
\begin{center}

\noindent{\Large \bf Bayesian total-evidence dating reveals the recent crown radiation of penguins} 

\bigskip



\noindent {\normalsize \sc Alexandra \ Gavryushkina$^{1,2}$, Tracy A.\ Heath$^3$, Daniel T.\ Ksepka$^4$, Tanja 
Stadler$^5$,  David Welch$^{1,2}$  and Alexei J.\ Drummond$^{1,2}$}\\
\noindent {\small \it 
$^1$Centre for Computational Evolution, University of Auckland, Auckland, New Zealand;\\
$^2$Department of Computer Science, University of Auckland, Auckland, 1010, New Zealand;\\
$^3$Department of Ecology, Evolution, \& Organismal Biology, Iowa State University, Ames, IA, 50011, USA;\\
$^4$Bruce Museum, Greenwich, CT, 06830, USA; \\
$^5$Department of Biosystems Science \& Engineering, Eidgen\"{o}ssische Technische Hochschule Z\"{u}rich, 4058 Basel, Switzerland}
\end{center}
\medskip
\noindent{\bf Corresponding authors:} Alexandra Gavryushkina and Alexei Drummond, Centre for Computational Evolution, University of 
Auckland, Auckland, New Zealand; E-mail: sasha.gavryushkina@auckland.ac.nz, alexei@cs.auckland.ac.nz\\


\vspace{1in}

\subsubsection{Abstract} The total-evidence approach to divergence-time dating uses molecular and morphological data from 
extant and fossil species  to infer phylogenetic relationships, species divergence times, and macroevolutionary parameters in a single 
coherent framework. Current model-based implementations of this approach lack an appropriate model for the 
tree describing the diversification and fossilization process and can produce estimates that lead to erroneous conclusions. We address this shortcoming by providing a total-evidence 
method implemented in a Bayesian framework.
This approach uses a mechanistic tree prior to describe the underlying diversification process that generated the tree of extant and fossil taxa. 
Previous attempts to apply the total-evidence approach have used tree priors that do not 
account for the possibility that fossil samples may be direct ancestors of other samples, that is, ancestors of fossil or extant species or of clades. The fossilized birth-death process explicitly
models the diversification, fossilization, and sampling processes and naturally allows for sampled ancestors. This model was recently applied to estimate divergence times based on 
molecular data and fossil occurrence dates. We incorporate the fossilized birth-death model and a model of 
morphological trait evolution into a Bayesian total-evidence approach to dating species phylogenies. We apply this method to extant 
and fossil penguins and show that the modern penguins radiated much more recently than
has been previously estimated, with the basal divergence in the crown clade occurring at $\sim$12.7 Ma
and most splits leading to extant species occurring in the last 2 million years. Our results demonstrate that including stem-fossil diversity can greatly improve 
the estimates of the divergence times of crown taxa. The method is available in BEAST2  (version 2.4) software \url{www.beast2.org} with 
packages SA (version at least 1.1.4) and morph-models (version at least 1.0.4) installed.   \\
\noindent{Keywords: phylogenetics, MCMC, calibration, divergence times, birth-death process} 


\vspace{1.5in}

Establishing the timing of evolutionary events is a major challenge in biology. 
Advances in molecular biology and computer science have enabled increasingly sophisticated methods for inferring phylogenetic trees. While the molecular data used to 
build these phylogenies are rich in information about the topological aspects 
of trees, these data only inform the relative timing of events in units of expected numbers of substitutions per site. The fossil record is  frequently used  to convert the timescale of inferred phylogenies to  absolute time \citep{zuckerkandl1962, zuckerkandl1965}.  
Exactly how to incorporate information from the fossil record into a phylogenetic analysis remains an active area of research.

Bayesian Markov chain Monte Carlo (MCMC) methods are now the major tool in phylogenetic inference
\citep{Yang1997,Mau1999,MrBayes}  and are implemented in  several widely used software packages \citep{PhyloBayes3,Beast17,MrBayes32,Beast2}.  
To date species divergences on an absolute time scale, Bayesian approaches must include three important components to decouple the confounded rate and time parameters: (1) a model describing how substitution rates are distributed across lineages, 
(2) a tree prior characterizing the distribution of speciation events over time and the tree topology, and (3) a way to incorporate information from the fossil or geological record to scale the relative times and rates to absolute values. 
Relaxed molecular clock models act as prior distributions on lineage-specific substitution rates and their introduction has greatly improved divergence-dating methods \citep{Thorne1998,drummond2006,Rannala2007,drummond2010,heath2012,LiDrummond2012}. 
These models do not assume a strict molecular clock, instead they allow each branch in the tree to have its own rate of molecular evolution drawn from a prior distribution of rates across branches. Stochastic branching models describing the diversification 
process that generated the tree are typically used as prior distributions for the tree topology and branching times \citep{yule1924,kendall1948,nee1994,rannala1996,Yang1997,gernhard2008,stadler2009}. 
When diversification models and relaxed-clock models are combined in a Bayesian analysis, it is possible to estimate divergence times on a relative time scale. External evidence, however, is needed to estimate absolute node ages. 

Various approaches have been developed to incorporate information from the fossil record or biogeographical dates into a Bayesian framework to calibrate divergence-time estimates  \citep{rannala2003, thorne2005, yang2006, Ho2009, heath2012b, Heled2012, parham2012, Silv2014, Heled2015}.
Calibration methods (also called `node dating') are the most widely used approaches for dating trees \citep{Ho2009} where absolute branch times are estimated using prior densities for the ages of a subset of divergences in the tree.
The placement of fossil-based calibration priors in the tree is ideally determined from prior phylogenetic analyses that include fossil and extant species, which could be based on analysis of morphological data alone, analysis of morphological data incorporating a backbone constraint topology based on molecular trees, or simultaneous analysis of combined morphological and molecular datasets. In practice, however,  fossil calibrations are often based on identifications of apomorphies in fossil material or simple morphological similarity. 

Node calibration using fossil constraints has two main drawbacks.
First, having  identified fossils as belonging to a clade,  a researcher needs to specify a prior distribution on the age of the common ancestor of the clade.  
Typically the oldest fossil in the clade is chosen as the minimum clade age but there is no agreed upon method of specifying the prior density beyond that. One way to specify a prior calibration density is through using the fossil 
sampling rate that can be estimated from fossil occurrence data \citep{footeRaup1996}.   
However, 
this approach must be executed with caution and attention to the quality of the fossil record for the clade of interest, as posterior estimates of divergence times are very sensitive to prior calibration densities of selected nodes 
\citep{warnock2012,dos2013,zhu2014,warnock2015} meaning that erroneous calibrations lead to erroneous results 
\citep{heads2012}. 

The second major concern about node calibration is that the fossilization process is modeled only indirectly and in isolation from other forms of data. A typical node-dating analysis is sequential: it first uses morphological 
data from fossil and extant species to identify the topological location of 
the fossils within a given extant species tree topology, then uses fossil ages to construct calibration densities, and finally uses molecular data to estimate the dated phylogeny.
Treating the different types of data in this sequential manner implies an independence between the processes that produce 
the different types of data, which
 is statistically inaccurate and errors at any step can propagate to subsequent analyses.
Furthermore, at the last step in the sequential analysis,  multiple different  
prior distributions are applied to estimate the dated phylogeny: a tree prior distribution and calibration distributions. 
Since these distributions all apply to the same object, they interact and careful consideration must be given to their specification so as to encode only the intended prior information \citep{Heled2012,warnock2015}. 
There is currently no efficient general method available to coherently specify standard tree priors jointly with calibration distributions \citep{Heled2015}.

In the total-evidence approach to dating \citep{lee2009, Pyron2011, Ronq2012}, one specifies a probabilistic model that encompasses the fossil data, molecular data and morphological data and then jointly estimates parameters of that model, 
including a dated phylogeny, in a single analysis using all available data.  It  builds on 
previously described methods for combining molecular and morphological data to infer phylogenies \citep{nylander2004} 
using a probabilistic model of trait evolution \citep[the Mk model of][]{Lewis2001}. The total-evidence approach to dating 
can be applied by employing a clock model and a tree prior distribution to calibrate the divergence times. The tree prior distribution 
describes the diversification process where fossil and extant species are treated as samples from this process.  
The placement of fossils and absolute branch times are determined in one joint inference rather than in separate  
analyses. The combination of clock models and substitution models for molecular and morphological data and a 
model of the process that generates dated phylogenetic trees with fossils 
comprises a full probabilistic model that 
generates all data used in the analysis.  

This approach can utilize all available fossils as individual data points. In contrast, the node calibration method only directly incorporates the age of the oldest fossil of a given clade, typically as a hard minimum for the clade age. The overall fossil record of the clade can be indirectly incorporated as the basis for choosing a hard or soft maximum or to justify the shape of a prior distribution, however individual fossils aside from the oldest will not contribute directly (except perhaps if they are used to generate a confidence interval).

Although total-evidence dating overcomes limitations of other methods that use fossil evidence to date 
phylogenies, some aspects of the method still need to be improved \citep{arcila2014}. One improvement is using 
better tree prior models. Previous attempts at total-evidence dating analyses have used  uniform, Yule, or birth-death tree priors 
that do not model the fossil sampling process and do not allow direct ancestors among the sample \citep[e.g.,][]{Pyron2011,Ronq2012,Wood2012}. 
However the probability of ancestor-descendant pairs among fossil and extant samples is not negligible \citep{Foote1996}. 
Moreover, ancestor-descendant pairs need to be considered when incomplete and non-identified specimens are included in the analyses 
because such specimens might belong to the same single lineages as other better preserved fossils. 

A good choice of the tree prior model is important for dating methods due to the limited amount of fossil data. 
\citet{dos2013} and \citet{zhu2014} showed that calibration methods are not statistically consistent, that is,  increasing the
amount of sequence data with a fixed number of calibration points does not decrease the uncertainty in divergence 
time estimates. \citet{zhu2014} conjectured that total-evidence approaches are not statistically consistent, 
implying that the speciation process assumptions play a significant role in dating phylogenies.

The fossilized birth-death (FBD) model \citep{Stadler2010, Didier12, StadKuhn12} explicitly models the fossilization process together 
with the diversification process and accounts for the possibility of sampled direct ancestors. \citet{Heath2014} 
used this model to estimate divergence times in a Bayesian framework from molecular data and fossil occurrence 
dates on a fixed tree topology. A comparison of different divergence dating methods showed that total-evidence 
analyses with simple tree prior models estimated significantly older divergence ages than analyses of molecular 
data and fossil occurrence dates with the FBD model \citep{arcila2014}.

Until recently, combining the FBD model with a total-evidence dating approach was complicated by the fact that existing 
implementations of the MCMC algorithm over tree space did not allow trees with sampled ancestors. 
\citet{gavr2014} addressed this problem and enabled full Bayesian inference using FBD 
model in the BEAST2 software \citep{Beast2} with the SA package (\url{https://github.com/CompEvol/sampled-ancestors}). This extended the \citet{Heath2014} method by allowing 
uncertainty in the tree topology of the extant species and placement of fossil taxa. 
Additionally, \citet{Zhang2015} implemented a variant of the FBD process that accounts for diversified taxon sampling and applied this to a total-evidence dating analysis of Hymenoptera \citep{Ronq2012}.
This study demonstrated the importance of modeling the sampling of extant taxa when considering species-rich groups \citep{hohna2011}. 

Here we implement total-evidence 
dating with the FBD model by including morphological data to jointly estimate divergence times and the topological relationships of fossil and living taxa.
We applied this method to a fossil-rich data set of extant and fossil penguins, comprising both molecular and morphological character data~
\citep{ksepka2012}. Our analyses yield dated phylogenies of living and 
fossil taxa in which most of the extinct species diversified before the origin of crown penguins, congruent with previous estimates of penguin relationships based on parsimony analyses \citep{ksepka2012}. 
Furthermore, our analyses uncover a significantly younger age for the most recent common ancestor (MRCA) of living penguins than previously estimated \citep{baker2006,brown2008,subramanian2013,jarvis2014,li2014}.  

\section{Materials and Methods}

\subsection{MCMC approach} 

We developed a Bayesian MCMC framework for analysis of morphological and molecular data to infer divergence 
dates and macroevolutionary parameters. The MCMC algorithm takes molecular sequence data from extant 
species, morphological data from extant and fossil species and fossil occurrence dates (or fossil occurrence 
intervals) as input data and simultaneously estimates dated species phylogenies (tree topology and divergence times), 
macroevolutionary parameters, and substitution and clock model parameters. We assume here that the gene phylogeny coincides with the species phylogeny. 
The state space of the Markov chain is a dated species phylogeny, $\mathcal T$, substitution and clock model parameters, $\bar 
\theta$, and tree prior parameters, $\bar \eta$. The posterior distribution is,
$$ f(\mathcal T, \bar \theta, \bar \eta | D, \bar \tau) \propto f(D, \bar \tau | \mathcal T, \bar \theta, \bar \eta) 
f(\mathcal T, \bar \theta, \bar \eta) =  f(D | \mathcal T, \bar \theta)  f(\bar \tau | \mathcal T) f(\mathcal T| \bar 
\eta)f(\bar \eta)f(\bar \theta),$$ 
where $D$ is a matrix of molecular and morphological data and $\bar \tau$ is a vector of time intervals assigned to 
fossil samples. On the right hand side of the equation, there is a tree likelihood function, $f( D | \mathcal T, \bar 
\theta)$,  a tree prior probability density, $f(\mathcal T | \bar \eta)$, prior probability densities for the parameters, 
and a probability density, $f(\bar \tau |  \mathcal T)$, of obtaining stratigraphic ranges $\bar \tau$, given $\mathcal T$
(remember, that $\mathcal T$ defines the exact fossilization dates). 
The tree prior density $f(\mathcal T | \bar \eta)$ is defined by 
equation (2) or (7) in \citet{gavr2014}. 

The full model describes the tree branching  process, morphological and molecular evolution along the tree, 
fossilization events, and assignment of the stratigraphic ranges to fossil samples, since we do not directly 
observe when a fossilization event happened. Thus the stratigraphic ranges for the fossils are considered as data.
We do not 
explicitly  model the process of the age range assignment but assume that for a fossilization event that happened 
at time $t$ the probability of assigning ranges $\tau_1>\tau_2$ does not depend on $t$ (as a function) if $\tau_1 > t \ge \tau_2$ 
and is zero otherwise. This implies that $f(\bar \tau |  \mathcal T)$ is a constant whenever the sampling times are 
within $\bar \tau$ intervals and zero otherwise and we get: 
\begin{equation}
\label{mcmc}
f(\mathcal T, \bar \theta, \bar \eta | D, \bar \tau) \propto f(D | \mathcal T, \bar \theta) \delta (\mathcal T \in T_{\bar 
\tau}) f(\mathcal T| \bar \eta)f(\bar \eta)f(\bar \theta),
\end{equation} 
where $T_{\bar \tau}$ is a set of phylogenies with sampling nodes within corresponding $\bar \tau$ intervals. 

\subsection{Modeling the speciation process}

We describe the speciation process with the fossilized birth-death model conditioning on sampling 
at least one extant individual \citep[equation 2 in][]{gavr2014}. This model assumes a constant rate birth-death 
process with birth rate $\lambda$ and death rate $\mu$ where fossils are sampled through time according 
to a Poisson process with a constant sampling rate $\psi$ and extant species are sampled at present with 
probability $\rho$. The process starts at some time  $t_{or}$ in the past --- the time of origin, where time is 
measured as a distance from the present. This process produces species trees with sampled two-degree nodes which we 
call sampled ancestors \citep[following][]{gavr2013,gavr2014}. Such nodes represent fossil samples and lie directly 
on branches in the tree. They are direct ancestors to at least one of the other fossils or extant taxa that has been sampled. 

We need to clarify what we mean by {\it sampling}. We have two types of sampling: fossil sampling and extant sampling. Suppose an individual from a population represented by a branch in the full species tree fossilized at some time in the past. Then this fossil was discovered, coded for characters and included in the analysis.  This would correspond to a fossil sampling event. Further, if an individual from one of the extant species was sequenced or recorded for morphological characters and these data are included to the analysis we say that an extant species is sampled. Suppose one sampled fossil belongs to a lineage that gave rise to a lineage from which another fossil or extant species was sampled. In such a case we obtain a sampled ancestor, that is, the former fossil is a sampled ancestor and the species to which it belongs is ancestral to the species from which the other fossil or extant species was sampled. If two fossils from the same taxon with different age estimates are included in an analysis, 
the older fossil has the potential to be recovered as a direct ancestor and would be considered
a sampled ancestor under our model.

In most cases,  we re-parameterize the fossilized birth-death model with $(t_{or}, d, \nu, s, \rho)$ where 
\begin{equation}
\label{dns} 
\begin{tabular}{ll} 
$d = \lambda-\mu$ & \text{\emph{net diversification rate}} \\ 
$\turnover = \frac \mu \lambda$ & \text{\emph{turnover rate}}\\ 
$\fosp = \frac\psi {\mu + \psi}$ & \text{\emph{fossil sampling proportion}} \\ 
\end{tabular} 
\end{equation}
These parameters are commonly used to describe diversification processes.  We also use the standard 
parameterization $(t_{or}, \lambda, \mu, \psi, \rho)$ assuming $\lambda > \mu$ in some analyses. Note, that the 
time of origin is a model parameter as opposed to the previous application of the FBD model \citep{Heath2014} 
where instead, the process was conditioned on the age of the MRCA, i.e., the oldest bifurcation node leading to 
the extant species, and all fossils were assumed to be descendants of that node. Here, we allow the 
oldest fossil to be the direct ancestor or sister lineage to all other samples because there is no prior evidence ruling those scenarios out.

\subsection{Bayes factors}

To assess whether there is a signal in the data for particular fossils to be sampled ancestors we calculated Bayes factors for each fossil. By definition a Bayes factor is: 
$$BF = \frac {P(D, \bar \tau|H_1, M)}{P(D, \bar \tau|H_2,M)} = \frac {P(H_1|D,\bar \tau, M)P(H_2|M)}{P(H_2|D,\bar \tau,M)P(H_1|M)},$$
where $H_1$ is the hypothesis that a fossil is a sampled ancestor, $H_2$ is the hypothesis that it is a terminal node, and $M$ is the combined model of speciation and morphological and molecular evolution. Thus $P(H_1|D,\bar \tau, M)$ is the posterior probability that a fossil is a sampled ancestor and $P(H_2|D,\bar \tau, M)$ that it is a terminal node and $P(H_1|M)$ and $P(H_2|M)$ are the corresponding prior probabilities. 

The Bayes factor reflects the evidence contained in the data for identifying a fossil as a sampled ancestor and compares the prior probability to be a sampled ancestor to the posterior probability. However we could not calculate the probabilities $P(H_1|M)$ and $P(H_2|M)$, so instead we looked at the evidence added by the morphological data towards identifying a fossil as a sampled ancestor to the evidence contained in the temporal data. That is, we replaced prior probabilities $P(H| M)$ with posterior probabilities given that we sampled 19 extant species and 36 fossils and assigned age ranges $\bar \tau$ to fossils, $P(H|\bar \tau, M)$, and calculated analogues of the Bayes factors:
$$\hat {BF} =  \frac {P(H_1|D,\bar \tau, M)P(H_2| \bar \tau, M)}{P(H_2|D,\bar \tau,M)P(H_1|\bar \tau, M)}.$$

To approximate $P(H|\bar \tau, M)$, we sampled from the distribution:  
\begin{equation}\label{condPrior}
f(\mathcal T, \bar \eta | \bar \tau) \propto \delta (\mathcal T \in T_{\bar 
\tau}) f(\mathcal T| \bar \eta)f(\bar \eta)
\end{equation} using MCMC. Having a sample from the posterior distribution~\eqref{mcmc} and  a sample from the conditioned prior distribution~\eqref{condPrior} we calculated $P(H_1|D, \bar \tau, M)$ and $P(H_1|\bar \tau, M)$ as fractions of sampled trees in which the fossil 
appears as a sampled ancestor in corresponding MCMC samples. Similarly,  we calculated $P(H_2|D, M; \bar \tau)$ and $P(H_2|M;\bar \tau)$ using trees in which the fossil is a terminal node. 

\subsection{Data}

We analysed a data set from \citet{ksepka2012} consisting of morphological data from fossil and living penguin 
species and molecular data from living penguins. The morphological data matrix used here samples 36 fossil species (we excluded 
{\it Anthropornis} sp.\,UCMP 321023 due to absence of the formal description for this specimen) and 19 extant species (we treated the Northern, Southern, and Eastern Rockhopper penguins as three distinct species for purpose of the analysis).
The original matrix contained 245 characters. We excluded outgroup taxa (Procellariiformes and Gaviiformes) because including them would violate the model assumptions: a uniform sampling of extant species is assumed whereas the matrix sampled all extant penguins but only a small proportion of outgroup species and also did not sample any fossil taxa from these outgroups. We excluded characters that became constant after excluding outgroup taxa, resulting in a matrix of 202 characters. 
The morphological characters included in the matrix ranged from two- to 
seven-state characters. The majority of these characters (\textgreater 95\%) have fewer than four states. 
Further, 48 of the binary characters were coded as present/absent. 
The molecular alignment comprises the nuclear recombination-activating gene 1 (RAG-1), and the mitochondrial 12S, 16S, cytochrome oxidase 1 (CO1), and cytochrome b genes. Each region is represented by more than 1,000 sites with 8,145 sites in 
total. Some regions are missing for a few taxa. 

The morphological dataset was originally developed to resolve the phylogenetic placement of fossil and extant penguins in a parsimony framework. Thus, efforts were focused on parsimony-informative characters. Though some apomorphic character states that are observed only in a single taxon are included in the dataset, no effort was made to document every possible autapomorphy. Thus such characters can be expected to be undersampled.  As with essentially all morphological phylogenetic datasets, invariant characters were not scored.

We updated the fossil stratigraphic ages---previously summarized in 
\citet{ksepka2010} --- to introduce time intervals for fossil samples as 
presented in online supplementary material (SM), Table~1. For fossil species known from a single specimen,
fossil stratigraphic ages represent the uncertainty related to the dating of the layer in 
which the fossil was found. For fossils known from multiple specimens, the ages
were derived from the ages of the oldest and youngest specimens. 

\subsection{Morphological evolution and model comparison}

We apply a simple substitution model for morphological character evolution --- the Lewis Mk model 
\citep{Lewis2001}, which assumes  a character can take $k$ states and the transition rates from one state to 
another are equal for all states. We do not model ordered characters and treated
34 characters that were ordered in the \citet{ksepka2012} matrix as unordered. 

Evolution of morphological characters has a different nature from evolution of 
DNA sequences and therefore requires different assumptions. In contrast to  molecular evolution 
models, we do not know the number of states each character can take and the number of states is not constant for different characters.
We consider two ways to approach this problem. First we can assume that the 
number of possible states for a character is equal to the number of different observed states. Typically, one would count
the number of different states in the data matrix for the character. Here, we obtained the number of observed states from the larger data matrix used in~\citet{ksepka2012} containing 13 outgroup species. Having the number of observed characters for each character, we partition the morphological data matrix in groups of characters having the same 
number of states and apply a distinct substitution model of the corresponding dimension to each partition. Another 
approach is to treat all the characters as evolving under the same model. The model dimension in this case is the 
maximum of the numbers of states observed for characters in the matrix. We refer to the first case as `partitioned 
model' and to the second case as `unpartitioned model'. Another difference comes from the fact that typically  
only variable characters are recorded. Thus the second adjustment to the model accounts for the fact that constant 
characters are never coded. This model is called the Mkv model \citep{Lewis2001}. 
We compared a model which assumed no variation in 
substitution rates of different morphological characters with a model using gamma distributed rates with a shared shape parameter for all 
partitions. We also compared a strict clock model and an uncorrelated relaxed clock model \citep{drummond2006} with a 
shared clock rate across partitions. To assess the impact of different parameterizations of the FBD model that induce slightly different prior distributions of trees we also considered two parameterizations ($d,\turnover,\fosp,\rho$--parameterization versus $\lambda,\mu,\psi,\rho$--parameterization).  

We completed a model selection analysis comparing different 
combinations of the assumptions for the Lewis Mk model, morphological substitution rates, and FBD model assumptions 
by running eight analyses of morphological data with different model settings. We then estimated the marginal likelihood 
of the model in each analysis using a path sampling algorithm \citep{baele2012}, with 20 steps and $\beta$-powers 
derived as quantiles of the Beta distribution with $\alpha_b=0.3$ and $\beta_b=1.0$. The traditional model 
selection tool is a Bayes factor, which is the ratio of the marginal likelihoods of two models: $M_1$ and 
$M_2$. A Bayes factor greater than one indicates that model $M_1$ is preferred over model $M_2$. Following 
this logic, the model that provides the best fit is the model with the largest marginal likelihood.  The model 
combinations with marginal likelihoods are described in Table~\ref{analyses}. 

\begin{table}[!h]
\caption{The tree-prior parameterization, clock, and substitution models used for eight analyses of penguin morphological data with marginal likelihoods for model testing.}
\begin{center}
\begin{minipage}{6in}
\begin{tabular}{c c c c c p{3.5cm}<\centering c}
\hline
\# & \shortstack[c]{Parti- \\ tions} & Gamma & \shortstack[c]{Lewis \\ model} & Clock & Parameterization &  
\shortstack[c]{Marginal \\ log-likelihood \\ from two runs} \\
\hline 
1 &  &  & Mk & Strict & $d, \turnover, \fosp$  & $-2644.21$, $-2644.16$  \\ 
2 &  & G &  Mk & Strict & $d, \turnover, \fosp$ & $-2641.62$, $-2642.94$ \\ 
3 & P &  & Mk & Strict & $d, \turnover, \fosp$ & $-1875.35$, $-1876.3$ \\ 
4 & P & G & Mk & Strict & $d, \turnover, \fosp$ & $-1859.84$, $-1860.37$\\ 
5 & P &  & Mk & Strict & $\lambda, \mu, \psi$ ($\mu < \lambda$) & $-1874.14$, $-1876.14$  \\
6 & P &  & Mkv & Strict & $d, \turnover, \fosp$ & $-1873.28$, $-1871.63$ \\ 
7 & P & G & Mkv & Strict & $d, \turnover, \fosp$ & $-1842.82$, $-1842.2$ \\ 
8\footnote{The analysis under model 8 has the largest marginal log-likelihood and was thus the model best supported by the data when evaluated using Bayes factors.} & P & G & Mkv & Relaxed & $d, \turnover, \fosp$ & $-1827.27$, $-1828.02$\\ 
\hline
\end{tabular}
\end{minipage}
\end{center}
\label{analyses}
\end{table}

\subsection{Posterior predictive analysis}

For most of the bifurcation events in trees from the posterior samples of the penguin analyses only one lineage 
survives while another lineage goes extinct. This suggests a non-neutrality in the evolution of the populations. To 
assess whether the FBD model, which assumes all lineages in the tree develop independently, fits the data 
analyzed here, we performed the posterior predictive analysis following \cite{drummond2008} under model 8 in 
Table~\ref{analyses}. The idea of such an analysis is to compare the posterior distribution of trees to the posterior 
predictive distribution \citep{gelman2013} and this type of Bayesian model checking has been recently developed 
for a range of phylogenetic approaches \citep{rodrigue2009,bollback2002,brown2014,lewis2014}.
The posterior predictive distribution can be approximated by the sample of trees simulated 
under parameter combinations drawn from the original posterior distribution. 
Out of computational convenience and similar to calculating $P(H|\bar \tau, M)$ for the Bayes factors, 
we conditioned the posterior predictive distribution on having sampled fossils within ranges, $\bar \tau$,
used in the original analysis. 

A way to compare posterior and posterior predictive distributions 
is to consider various tree statistics and calculate the tail-area probabilities by calculating the $p_B$ probability, which 
is simply the proportion of times when a given test statistic for the simulated tree exceeds the same statistic for 
the tree from the posterior distribution corresponding to the same parameters. Extreme values of $p_B$, i.\,e., 
values that are less than 0.05 or greater than 0.95, would indicate the data favor a non-neutral scenario. For 
this analysis, we considered tree statistics which can be grouped into two categories: statistics that describe the 
branch length distribution and the tree shape. The test statistics and corresponding \nohyphens{$p_B$-values} 
are summarized in Table~\ref{statistics}.

\begin{table}[!h]
\caption{The posterior-predictive analysis of the penguin data for branch-length and tree-imbalance statistics indicating no significant difference in the posterior and posterior predictive tree distributions for model 2 (in Table~\ref{analyses}).}  
\begin{center}
\begin{minipage}{5.8in}
{\footnotesize
\begin{tabular}{p{0.7\textwidth}lll}
\hline
Description & Notation & $p_B$-value\footnote{A $p_B$ value is the proportion of times a given test statistic for the simulated tree exceeds the value of that statistic for the tree from the posterior distribution.}\footnote{All $p_B$ values are within [0.05, 0.95]} \\
\hline
\multicolumn{3}{c}{Branch length distribution statistics} \\
\hline
The total length of all branches in the tree & $T$ & 0.83 \\
The ratio of the length of the subtree induced by extant taxa and the total tree length & $\rho^T_{trunk}$ & 0.33 \\ 
Genealogical Fu \& Li's $D$ calculated as the normalized difference between external branch length in the tree 
with suppressed sampled ancestor nodes and total tree length. & $D_F$ & 0.5 \\
The time of the MRCA of all taxa & $t_{MRCA}$  & 0.57 \\
The time of the MRCA of all extant taxa & $t_{EMRCA}$ & 0.46 \\
\hline 
\multicolumn{3}{c}{Tree imbalance statistics} \\
\hline
The maximum number of bifurcation nodes between a bifurcation node and the leaves summed over all 
bifurcation nodes except for the root & $B_1$ & 0.75 \\ 
Coless's tree imbalance index calculated as the difference between the numbers of leaves on two sides of a node 
summed over all internal bifurcation nodes and divided by the total number of leaves. & $I_c$ & 0.28 \\
The number of cherries (two terminal nodes forming a monophyletic clade, sampled ancestors are suppressed) & $C_n$ & 0.54 \\
The number of sampled ancestors & SA  & 0.26 \\
\hline
\end{tabular}
}
\end{minipage}
\end{center}
\label{statistics}
\end{table} 

\subsection{Total-evidence analysis of penguins}

For the total-evidence analysis of the penguin data set we chose the substitution and clock models for 
morphological character evolution with the largest marginal likelihood from the model comparison analysis 
(analysis 8 in Table~\ref{analyses}). This model suggests that the data are partitioned in groups of characters with respect to the number 
of observed states. Each partition evolves under the Lewis Mkv model and the substitution rate varies across 
characters according to a Gamma distribution shared by all partitions. The morphological clock is modeled with 
an uncorrelated relaxed clock model with log-normal distributed rates. For molecular data we assume a general-time reversible model with gamma-distributed rate heterogeneity among sites (GTR+$\Gamma$) 
for each of the five loci with separate rate, frequency, and gamma shape parameters for each 
partition. A separate log-normal uncorrelated relaxed clock model is assumed for the molecular data alignment. 
Each branch is assigned a total clock rate drawn from a log-normal distribution and this rate is scaled by a relative 
clock rate for each gene so that relative clock rates for five partitions sum up to one. We also ran the same analysis 
under the strict molecular clock. We assumed the FBD model as a prior 
distribution for time trees with uniform prior distributions for turnover rate and fossil sampling proportion, log-normal prior 
distribution for net diversification rate with 95\% highest probability density (HPD) interval covering $[0.01,0.15]$ 
estimated in \citep{Jetz2012} and sampling at present probability fixed to one since all modern penguins were included in the analyses. 
We also analyzed this data set under the birth-death model without sampled ancestors~\citep{Stadler2010}. 
 
In addition to analyzing the full data set, we performed a separate analysis of only living penguins and the crown fossils, to examine the effect of
ignoring the diversification of fossil taxa along the stem lineage. Crown fossils were selected if the fossil lineage was a descendant of the MRCA of all extant species with a posterior probability 
greater than 0.05 in the full analysis (i.e., the analysis with stem and crown fossils). Thus, this analysis includes 6 fossils 
(listed in Table~\ref{crownFossils}) and all living penguins. 
For this analysis we did not condition on recording only variable 
characters (i.e., we used the Mk model) because after removing a large proportion of stem fossils, some characters become constant.

\renewcommand*\footnoterule{}
\renewcommand{\thempfootnote}{\arabic{mpfootnote}}
\begin{table}[!h]
\caption{The posterior probability of a fossil's placement in the crown (only for fossils with non-zero probability).} 
\begin{center}
\begin{minipage}{3.25in}
\begin{tabular}{l c}
\hline
Fossil\footnote{The six fossils with probabilities greater than 0.05 were used for the total-evidence analysis without stem fossils.} & Probability \\ 
\hline
{\it Spheniscus megaramphus} & 0.9992 \\     
{\it Spheniscus urbinai} & 0.9991 \\ 
{\it Pygoscelis grandis} & 0.9928 \\ 
{\it Spheniscus muizoni} & 0.9201 \\ 
{\it Madrynornis mirandus} & 0.9007 \\ 
{\it Marplesornis novaezealandiae} & 0.1652 \\ 
{\it Palaeospheniscus bergi} &  0.0001 \\   
{\it Palaeospheniscus biloculata} & 0.0001 \\  
{\it Palaeospheniscus patagonicus} & 0.0001 \\   
{\it Eretiscus tonnii} & 0.0001 \\
\hline     
\end{tabular}
\end{minipage}
\end{center}
\label{crownFossils}
\end{table}

\newpage

\subsection{Summarizing trees} First we summarized the posterior distribution of full trees using summary 
methods from \cite{heled2013}. As a summary tree we 
used the maximum sampled-ancestor clade credibility tree. 
A maximum sampled-ancestor clade credibility tree is a tree from the posterior sample that maximizes the product of posterior clade probabilities. 
Here, a clade denotes two types of objects. The first type is a monophyletic set of taxa with a bifurcation node as the MRCA. 
Such clades are completely defined by a set of taxon labels $\{B_1, \ldots, B_n\}$ meaning that we do not distinguish between clades with the same 
taxon set but different topologies. The second type is a monophyletic set of taxa with a two-degree sampled node
as the MRCA. This can happen when one of the taxa in the group is a sampled ancestor and 
it is ancestral to all the others in the clade. Then this taxon will be the MRCA of the whole group assuming it is an ancestor to itself. These clades are defined by the pair $(B_i, \{B_1, \ldots, B_n\})$ where $\{B_1, \ldots, B_n\}$ are taxon labels and $B_i$, $1 \leq i \leq n$, is the taxon that is ancestral to all taxa in the clade.

Second, we removed all fossil lineages from the posterior trees thereby suppressing two-degree nodes and then summarized the resulting 
trees (which are strictly bifurcating) with a maximum clade credibility tree with common ancestor ages.
To assign a common ancestor age to a clade, we consider a set of taxa contained in the clade and find the age of the MRCAs of 
these taxa in every posterior tree (including the trees where these taxa are not monophyletic) and take the 
mean of these ages.

\newpage 

\section{Results and Discussion}

\subsection{Model comparison}

\begin{figure}
\caption{Posterior probabilities of fossils to be sampled ancestors for eight models summarized in Table~
\ref{analyses}. 
In the legend, P stands for the partitioned model, G for gamma variation across sites, Mkv for conditioning on 
variable characters, R for relaxed clock model, dns for $d$, $\nu$, and $s$ tree prior parameterization, lmp for $
\lambda$, $\mu$, and $\psi$ tree prior parameterization and the numbers correspond to analyses in Table~\ref{analyses}. }
\includegraphics[height=0.65\textheight]{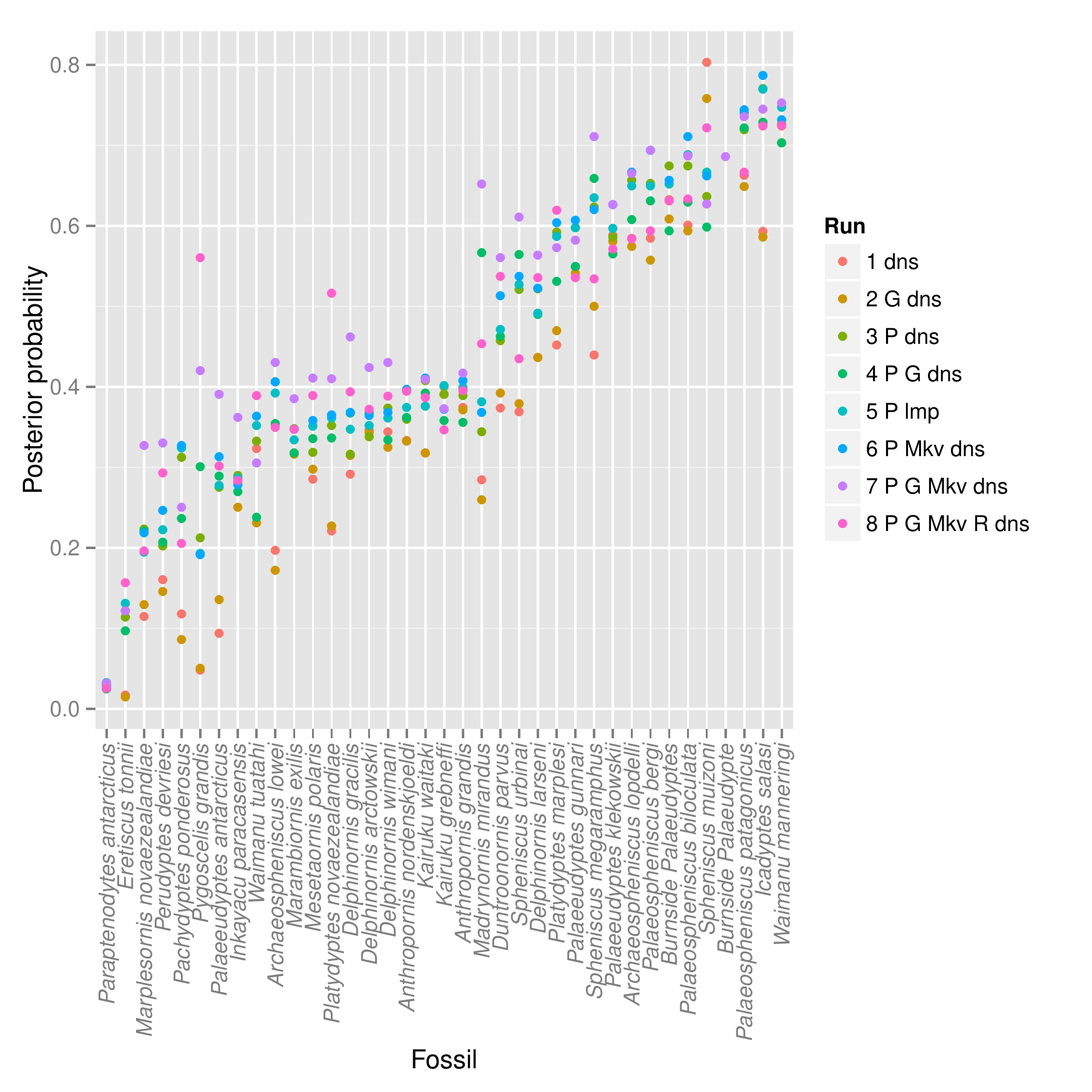}
\label{saPostPr}
\end{figure}

For each of the eight analyses listed in Table~\ref{analyses}, we plotted the probability of each fossil to be a 
sampled ancestor (Figure~\ref{saPostPr}). This shows that assumptions about the clock and substitution 
models as well as the tree prior model contribute to identifying a fossil as a sampled ancestor. The comparison of 
the marginal likelihoods for different assumptions about the clock and substitution models shows that the 
substitution model where characters are partitioned in groups with the same number of states and with gamma 
variation in the substitution rate across characters is the best model for this dataset. 

\subsection{Posterior predictive analysis}

\begin{figure}
\caption{The posterior (red) and posterior predictive (blue) distributions for the tree length, $T$, and genealogical $D_F$ statistics on 
the left and $B_1$ tree imbalance statistic and Colless's tree imbalance index, $I_c$, on the right for model 8 in 
Table~\ref{analyses}. The plots do not show obvious discrepancy in the posterior and posterior predictive distributions of 
these statistics. The posterior predictive distribution for the branch length related statistics ($D_F$ and $T$) is more diffuse 
than the posterior distribution although both distributions concentrate around the same area. The distributions of the tree imbalance
statistics almost coincide.} 
\includegraphics[width=0.9\textwidth]{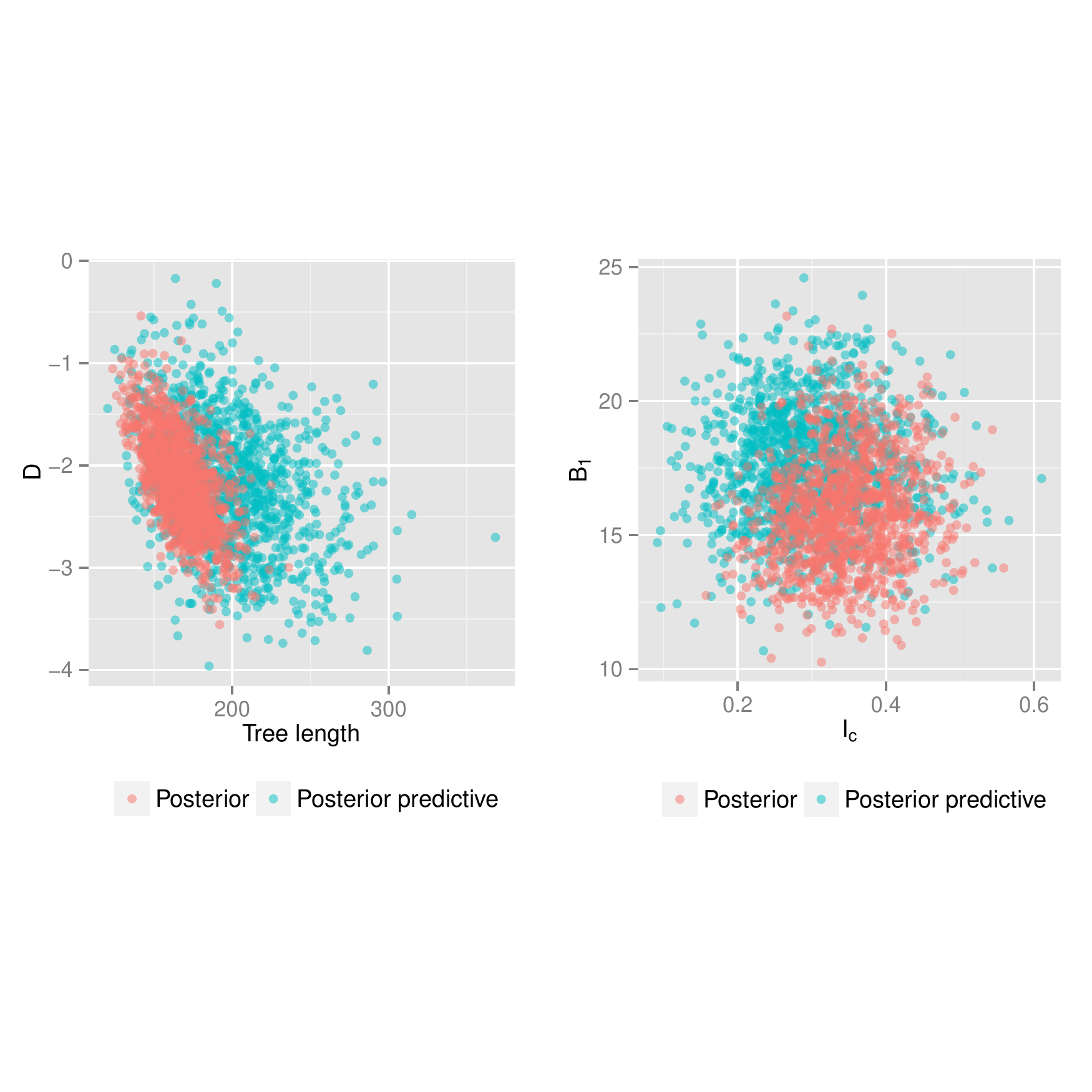}
\label{postPredictive}
\end{figure}

The posterior predictive analysis did not reject the FBD process, where lineages evolve independently of
each other, as an adequate model for describing 
the speciation-extinction-fossilization-sampling process for these data. 
The $p_B$ values for all nine statistics were within the $[0.26, 0.83]$ interval (Table~\ref{statistics}). 
The plots of the posterior and posterior predictive distributions for several statistics (Fig.\,\ref{postPredictive}) show 
that there is no obvious discrepancy in these distributions. 
Thus there is no signal in the data to reject a neutral diversification of penguins. 

In this analysis, we conditioned the posterior predictive distribution to have the given age ranges. To assess 
the overall fit of the FBD model, one needs to perform a posterior predictive analysis where the posterior predictive distribution is not conditioned on the age ranges nor on the number of sampled nodes. We have not performed such an analysis.

\subsection{Penguin phylogeny}   

\begin{landscape}

\begin{figure}
\begin{center}
\caption{\footnotesize A maximum sampled ancestor clade credibility tree for the total-evidence analysis.  
The numbers in blue (at the bases of clades) show the posterior probabilities of the clades. The filled red and black circles represent sampled 
ancestors. Fossils with positive evidence of being sampled ancestors are shown in red. Fossils
{\it Paraptenodytes antarcticus} and {\it Palaeospheniscus patagonicus} both appear around the same time and have 
the same prior probabilities of 0.42 of being sampled ancestors but the morphological data provides 
positive evidence for the former to belong to a terminal lineage and for the latter to be a sampled ancestor. 
Penguin reconstructions used with permission from the artists: fossil species by Stephanie Abramowicz and extant penguin species by Barbara Harmon.}
\includegraphics[width=20cm]{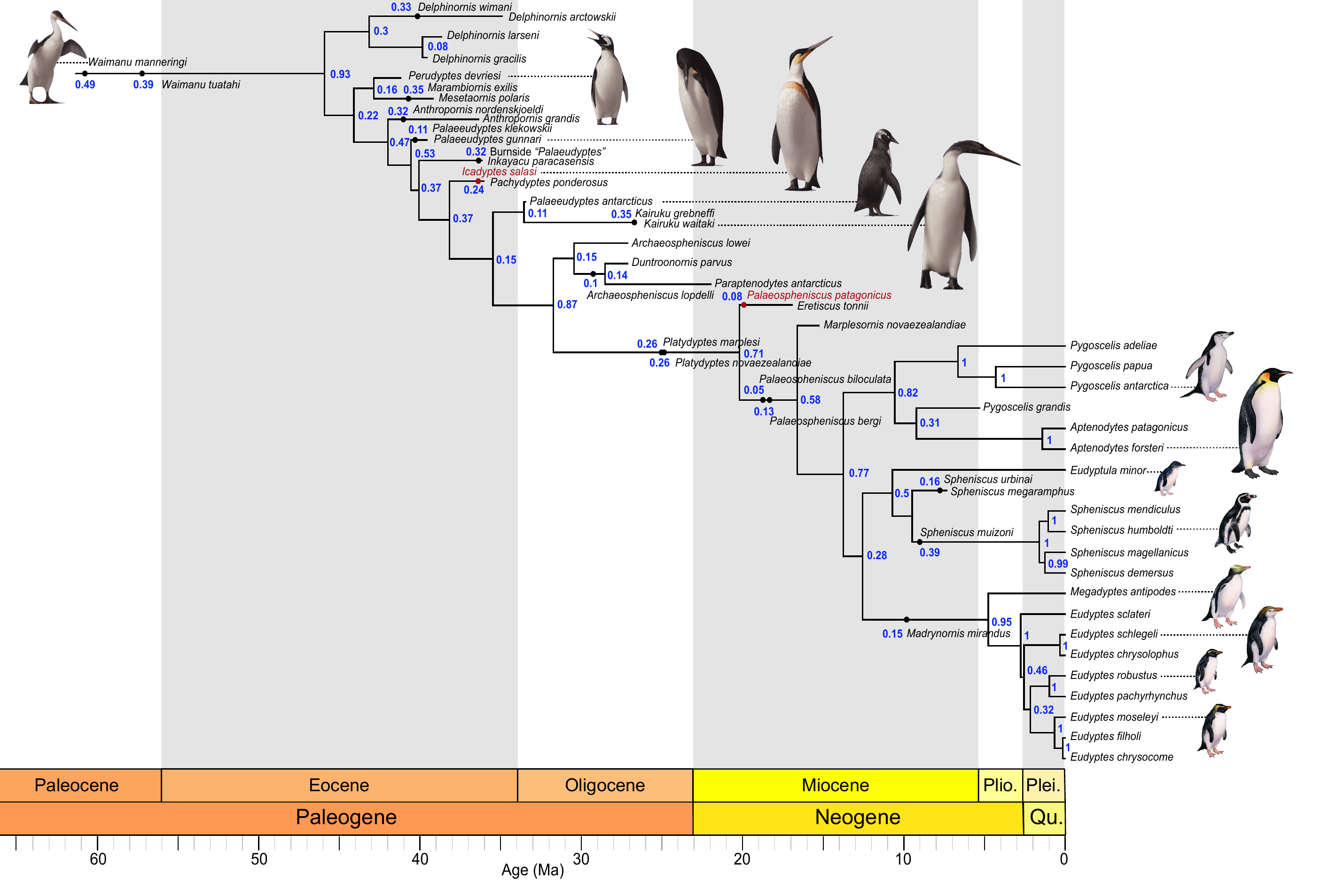}
\label{penguinTree}
\end{center}
\end{figure}

\end{landscape}

The maximum sampled-ancestor clade credibility tree (MSACC tree)  (Fig.\,\ref{penguinTree}) shows that most of the penguin 
fossils do not belong to the crown clade and that the crown clade Spheniscidae originated only $\sim$ 12.7 million years ago.
The posterior probabilities of most clades including fossils are low, reflecting the 
large uncertainty in the topological placement of the fossil taxa, 
whereas many clades uniting extant taxa receive substantially higher posterior probabilities.

\begin{figure}
\begin{center}
\caption{The evidence for fossils to be sampled ancestors. The samples above the shaded area (that is, with log Bayes Factors greater than one) have positive evidence to be sampled ancestors and below the shaded area (log Bayes 
factors lower than minus one) 
have positive evidence to be terminal nodes.}
\includegraphics[width=0.7\textwidth]{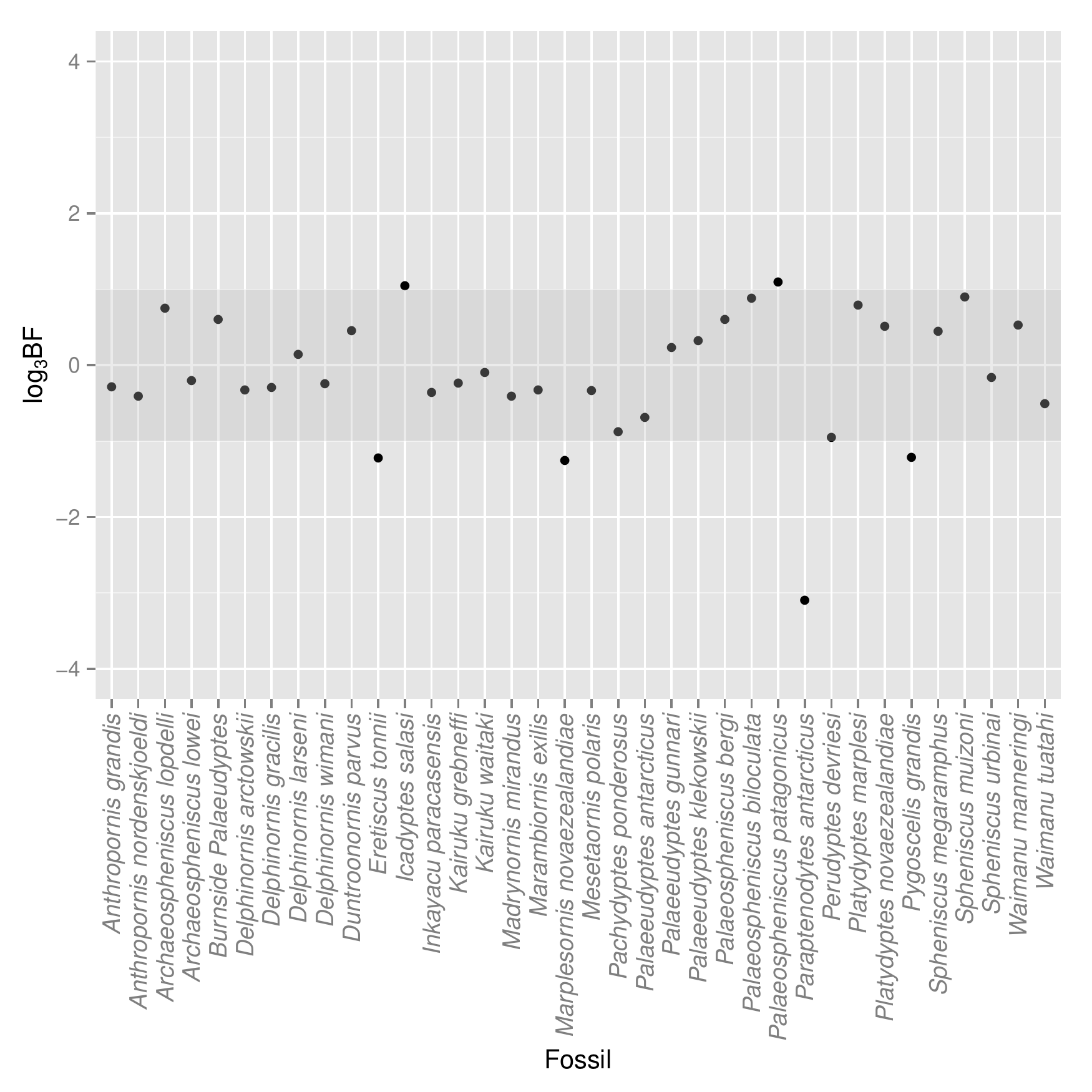}
\label{fig:BF}
\end{center}
\end{figure}

We calculated Bayes factors for all fossils to be sampled ancestors assuming the 
prior probability that a fossil is a sampled ancestor is defined by the tree prior model
conditioned on the number of sampled extant and fossil species and assigned sampling
intervals. Adding comparative (morphological and molecular) data to the sample size and
sampling intervals provides positive evidence that the fossil taxa representing the species {\it Palaeospheniscus patagonicus} and
{\it Icadyptes salasi} are sampled ancestors. 
{\it Eretiscus tonnii}, {\it Marplesornis novaezealandiae}, {\it Paraptenodytes antarcticus} and {\it Pygoscelis grandis} show positive evidence to be terminal samples (Fig.\,\ref{fig:BF}). 

Due to the large uncertainty in the topological placement of fossil 
taxa, the relationships displayed in the summary tree are not the only ones supported by the posterior distribution. Thus, in some cases an alternate topology cannot be statistically 
rejected and a careful review of the entire population of sampled trees is required to fully account for this.
Below, we summarize the
features of the MSACC topology that differ from previous estimates of penguin phylogeny, keeping this uncertainty in mind. 

\begin{figure}
\caption{The maximum clade credibility tree of extant penguins with common ancestor ages. The blue bars are 
the 95\% highest probability density (HPD) intervals for the divergence times. The mean estimates and 95\% HPD 
intervals are summarized in SM, Table~4. The numbers in blue (at the bases of clades) show the 
posterior probabilities of the clades (after removing fossil samples).}
\includegraphics[width=\textwidth]{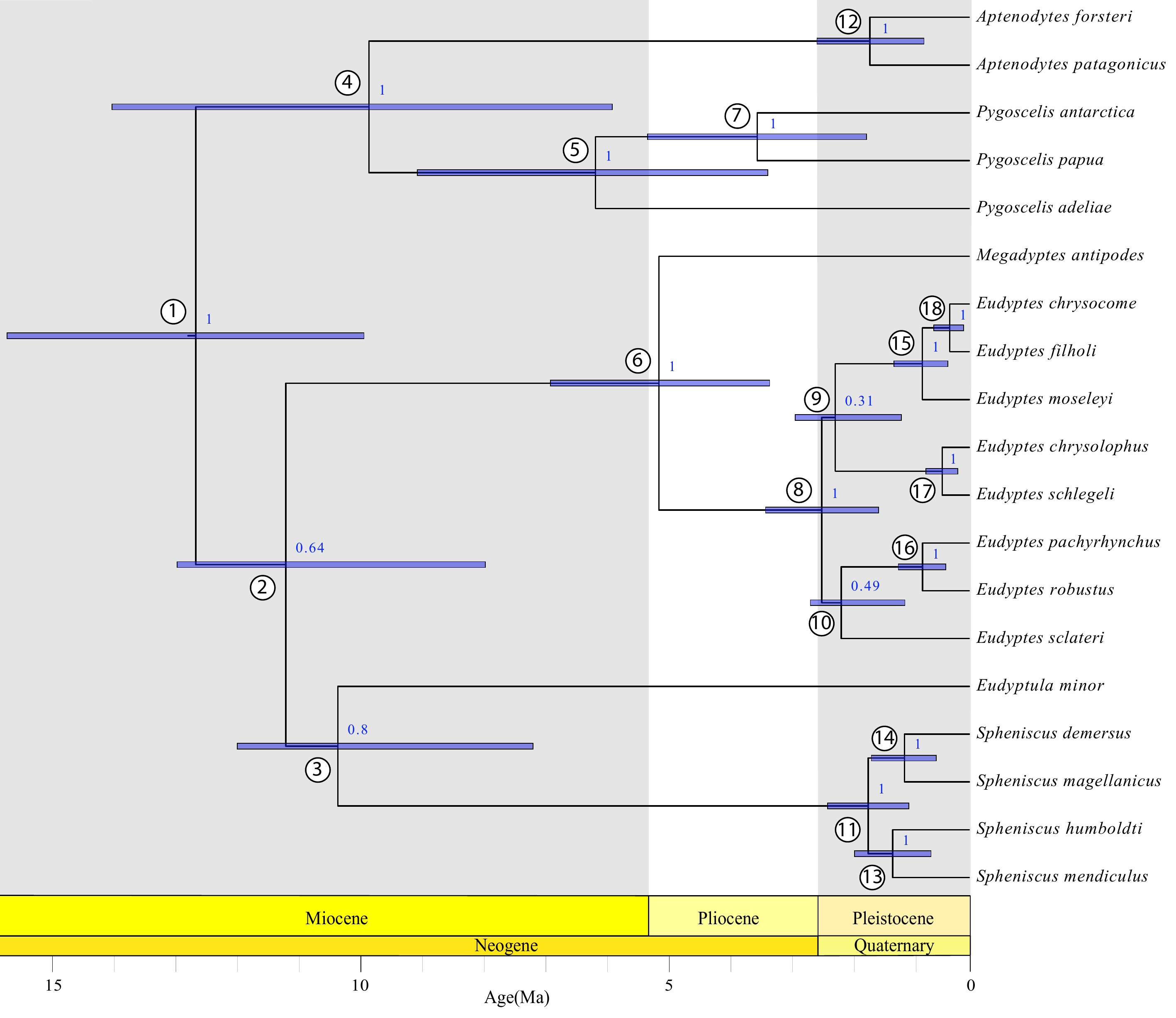}
\label{fig:divDates}
\end{figure}

The relationships within each genus are similar to those reported in previous parsimony analyses of the dataset \citep{ksepka2012}, with some exceptions within the crested penguin genus {\it Eudyptes}. These agree with the results of \citet{baker2006} based on Bayesian analysis of the same molecular loci (without morphological data), though it should be noted that our Bayesian analysis shows a degree of uncertainty in the resolution of the {\it Eudyptes} clade. 
The summary tree obtained after removing fossil taxa displays a different set of relationships within {\it Eudyptes} with high posterior probabilities (Fig.\,\ref{fig:divDates}). 

Allowing fossils to represent ancestors yields several interesting results. Although there is no evidence in comparative data to support an ancestral position for {\it Spheniscus muizoni} (Fig.\,\ref{fig:BF}) the combined comparative and temporal data yields the posterior probability of 0.61 that it is an ancestor of the extant {\it Spheniscus} radiation (that is, in 61\% of the posterior trees, this taxon is a direct ancestor of the four extant {\it Spheniscus} species and possibly some other species as well). The ancestral position is consistent with the morphological data set: {\it S. muizoni} preserves a mix of derived character states that support 
placement within the {\it Spheniscus} clade and primitive characters which suggests it falls outside the clade formed by the 
four extant {\it Spheniscus} species. Furthermore, at least for the discrete characters sampled, it does not exhibit apomorphies providing direct evidence 
against ancestral status.  

{\it Madrynornis mirandus} is recovered as ancestor to the {\it Eudyptes} + {\it Megadyptes} clade, though this placement receives low posterior probability (0.15). This fossil taxon was inferred as the sister taxon to {\it Eudyptes} by several previous studies \citep{hospitaleche2007,ksepka2010, ksepka2012}; though see \cite{chavez2014}, and so had 
been recommended as a calibration point for the {\it Eudyptes-Megadyptes} divergence \citep{ksepka2010} and used as such  \citep{subramanian2013}.  In our analysis,  a {\it Megadyptes} + {\it Eudyptes} clade excluding {\it Madrynornis} is present in all posterior trees, that is, the results reject the possibility that this taxon is the sister taxon to {\it Eudyptes} and its use as a calibration point is in need of further scrutiny. Our results indicate a 0.9 posterior probability that {\it Madrynornis mirandus} belongs in the crown, but do not provide solid support for the precise placement of this fossil taxon. Presumably the position of {\it Madrynornis mirandus} outside of {\it Eudyptes} + {\it Megadyptes} clade is at least partially attributable to the temporal data --- its age means a more basal position is more consistent with the rest of the data.

Most of the clades along the stem receive very low posterior probabilities, which is not unexpected given that some stem penguin taxa remain poorly known, with many morphological characters unscorable. These clades correspond to the large polytomies from~\citet{ksepka2012} (polytomies are not allowed in the summary method we used here). Of particular note is 
 the placement of {\it Palaeospheniscus bergi} and {\it Palaeospheniscus biloculata} on a branch along the backbone 
 of the tree leading to all modern penguin species. {\it Palaeospheniscus} penguins share many synapomorphies with crown penguins and only a single feature 
 in the matrix (presence of only the lateral proximal vascular foramen of the tarsometatarsus) contradicts this possibility \citep[and supports their forming a clade with {\it Eretiscus tonnii} in the strict consensus of][]{ksepka2012}.
 The posterior distribution of our analysis supports a clade containing the three {\it Palaeospheniscus} species and {\it Eretiscus tonnii} with probability 0.06,
and so this relationship cannot be ruled out.
 
 Overall, the estimated clades with large posterior probabilities (greater than 0.5)
agree with those clades previously estimated from the same 
 data set using parsimony methods \citep{ksepka2012}.  Low posterior probability values are due to the sparse morphological data 
 (and complete lack of molecular data) for many early stem taxa. Several species such as {\it Palaeeudyptes antarcticus} are 
 based on very incomplete fossils, in some cases a single element, and so we view the relationships estimated for deep stem 
 penguins as incompletely established for the time being. Despite the high degree of uncertainty in the phylogenetic relationships of fossil species, the overall support for the general scenario placing most fossil penguins along the stem with a recent appearance 
of crown penguins is strong. To better describe, measure, and visualize the topological uncertainty of total-evidence analyses, methods similar to \citet{billera2001}, \citet{owen2011}, and \citet{gavrAN2014} should be developed for serially sampled trees with sampled ancestors.

\subsection{Divergence dates}

We estimated the divergence dates for extant penguins (Fig.\,\ref{fig:divDates} and SM, Table~4). 
In general, the estimates are younger than those reported in previous studies: \citep{baker2006, brown2008, subramanian2013,jarvis2014,li2014}. \citet{baker2006} used the penalized-likelihood approach \citep{sanderson2002} with secondary calibrations, and estimated 
the origin of crown penguins to be 40.5 Ma  (95\% confidence interval: 34.2--47.6 Ma). 
\citet[][Fig.\,4]{brown2008} estimated this age at $\sim$50 Ma using a Bayesian approach with uncorrelated rates 
and 20 calibrations distributed through Aves, including the stem penguin {\it Waimanu manneringi}.
 \citet{subramanian2013} estimated a much younger crown age by using a Bayesian analysis with node calibration densities based on four fossil penguin taxa: {\it Waimanu manneringi}, {\it Madrynornis mirandus}, {\it Spheniscus muizoni}, 
and {\it Pygoscelis grandis}. Their estimate of the age of the MRCA of the extant penguins was 20.4 Ma (95\% HPD interval: 17--23.8 Ma)  \citep{subramanian2013}. 
Most recently, \citep{jarvis2014, li2014} estimated the age of the crown
penguins at 23 Ma (95\% confidence interval: 6.9--42.8 Ma) using a Bayesian method in MCMCTree~\citep{dos2011} based on genomes from two penguin species ({\it Aptenodytes forsteri} and {\it Pygoscelis adeliae}) and calibrations for higher avian clades including {\it Waimanu manneringi}. Notably, this last date can be considered applicable to the crown only if 
{\it Aptenodytes} or {\it Pygoscelis} is the sister taxon to all other penguins, otherwise 
this date applies to a more nested node, implying an older age for the crown.  

Our total-evidence analysis under the FBD model suggests that the MRCA is younger than any of these previous 
estimates at 12.7 Ma (95\% HPD interval [9.9,15.7]; see Fig.\,\ref{fig:divDates} and SM, Table~4). 
We assert that this is the best constrained 
estimate of the age of the penguin 
crown clade to date, because it avoids potential pitfalls related to the use of secondary calibrations, samples all extant species, 
and most importantly includes all reasonably complete fossil taxa directly as terminals or sampled ancestors. This final point is crucial, not only because 
including fossils as terminals has been shown influence phylogenetic accuracy under many conditions 
\citep[e.g.,][]{hermsen2008,grande2010,hsiang2015}, but also because at least one fossil taxon---previously used as a node calibration---was recovered at a more basal position in our results. The small gap between our 12.7 Ma 
estimate for the MRCA of extant penguins and the oldest identified crown fossil at $\sim$10 Ma is consistent with the fossil record of penguins as a whole, 
which includes a dense sampling of stem species from $\sim$60 Ma to $\sim$10 Ma, and only crown fossils from $\sim$10 Ma 
onwards. Moreover, our results suggest many extant penguin species are the product of recent divergence events, with 13 of 19 sampled species splitting from their sister taxon in the last 2 million years. 
Penguins have a relatively dense fossil record compared to other avian clades, with thousands of specimens known 
from four continents and spanning nearly the entirety of their modern day geographical range.  If crown penguins originated at 
20--50 Ma as implied by previous studies, it would require major ghost lineages \citep{clarke2007,clarke2014}, and thus a modest to extreme fossilization bias favoring the preservation of stem 
penguin fossils and disfavoring the preservation of crown penguin fossils. Such a bias is difficult to envision, as both stem and crown 
penguins share a dense bone structure and preference for marine habitats that would suggest similar fossilization potential. 

Inclusion of stem fossil diversity has a profound impact on the inferred age of the penguin crown clade. To demonstrate this effect, we performed 
a total-evidence analysis including only living penguins and crown fossils (i.e., fossil taxa identified as crown 
penguins in our primarily analysis). Both the age estimate and the inferred uncertainty in the MRCA age of crown penguins increased substantially with the MRCA age shifting to 22.8 Ma (95\% HPD interval: 14.2--33.6 Ma; Fig.\,\ref{fig:divDates}). This shows that including the stem-fossil diversity allows for a 
better estimate of the crown age of penguins --- one that is more consistent with the fossil record. Furthermore, 
these additional data points contribute to better estimates of diversification parameters.

For the complete analysis of stem and crown taxa, the mean estimate of the net diversification rate, $d$,
was 0.039 with [0.002, 0.089] HPD interval although this estimate is sensitive to the prior distribution (SM, Fig.\,2), 
the turnover rate, $\turnover$, was 0.88 ([0.72, 1]) and the sampling proportion, $s$, was 0.23 ([0.06, 0.43]). 

\begin{figure}
\begin{center}
\caption{The ages of the most recent common ancestors of the extant penguin taxa for the eight analyses (Table~\ref{analyses}) of 
morphological data, total-evidence analysis with all fossils under relaxed (8+DNA R) and strict (8+DNA S) molecular clocks, total-evidence analysis under the birth-death model without sampled ancestors (8BD+DNA R) and 
total-evidence analysis with crown fossils only. Abbreviations in the names of analyses are the same as in Figure~\ref{saPostPr}}
\includegraphics[width=0.6\textwidth]{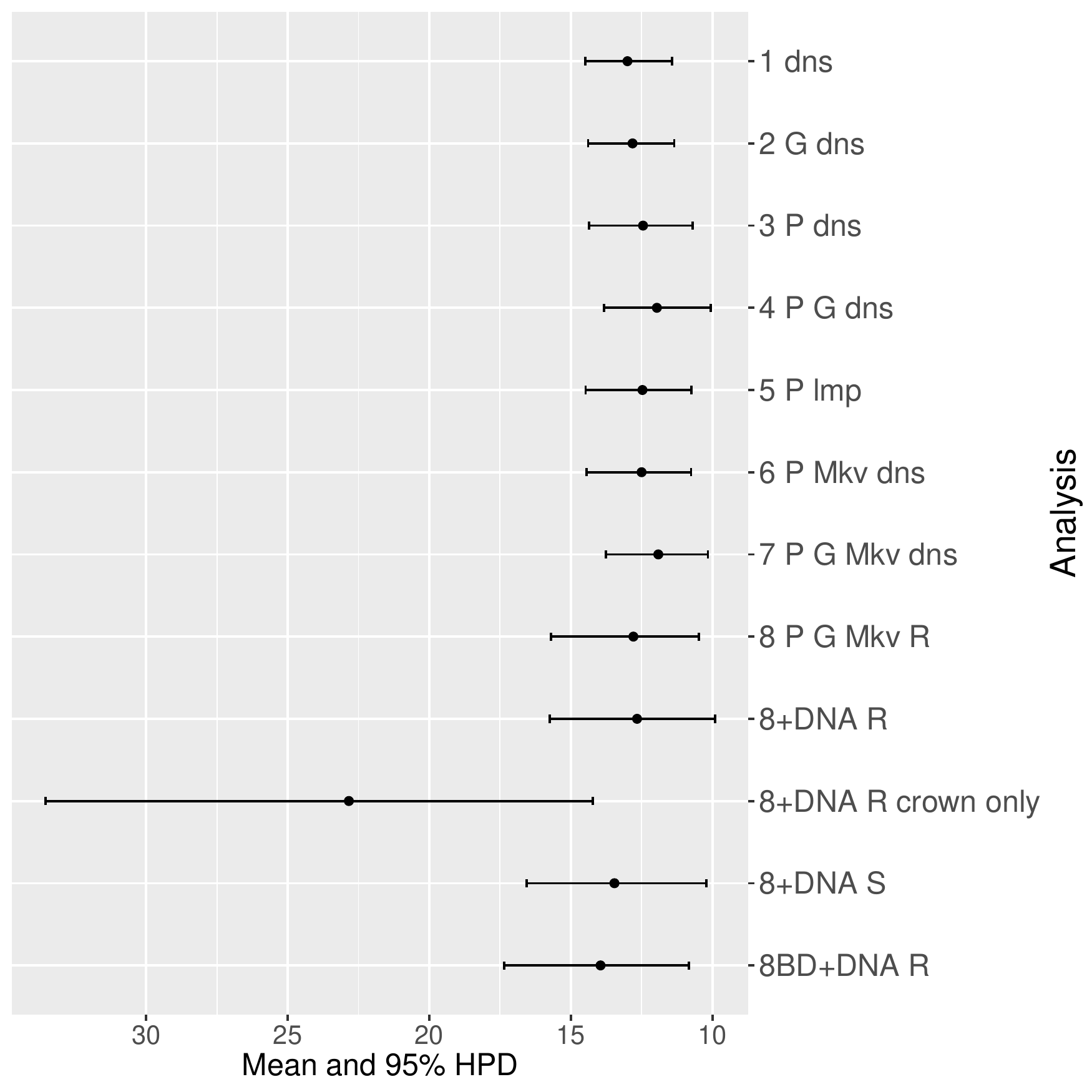}
\label{fig:rootAges}
\end{center}
\end{figure}

The posterior distribution for the scale parameter of the log-normal distribution in the uncorrelated relaxed 
clock model for the molecular alignment was bimodal with a mode around 0.4 and a mode around zero. 
This suggested a strict molecular clock model might fit the data. The additional analysis of this 
data set under the strict molecular clock slightly shifted the estimates of the time of the MRCA of crown penguins toward 
the past although the posterior intervals largely overlap (Fig.\,\ref{fig:rootAges}).

Using the birth-death-sampling model without sampled ancestors~\citep{Stadler2010} instead of the FBD model shifted the estimated divergence times 
toward the past (Fig.\,\ref{fig:rootAges} and SM, Fig.\,1). \citet{gavr2014} showed in simulation studies that ignoring sampled ancestors results in an increase in the diversification rate. We can observe the same trend here: the mean of the net diversification rate for the analysis under the birth-death model without sampled ancestors was 0.092 (95\% HPD: [0.007, 0192]) compared to 0.039 ([0.002, 0.089]) for the FBD model. Although we would expect a decrease in the diversification rate in older trees on the same number of extant tips, in the birth-death model without sampled ancestors, a sampling event causes an extinction of the lineage. Thus, the diversification rate (here, modeled as the difference in the birth and death rates) does not account for the extinction `by sampling'. The mean estimate for $(\lambda - \mu - \psi)$, which better describes the diversification rate in the birth-death model without sampled ancestors, was 0.019 ([-0.061,0.099]). 
 
\subsection{Implications for crown penguin evolution}

With many extant penguin species inhabiting extreme polar environments,
penguin evolution is often considered through the lens of global climate change. The fossil record has revealed that, despite their 
celebrated success in modern polar environments, penguins originated during a warm period in Earth's history, and the first 
Antarctic penguins were stem taxa that were distantly related to extant Antarctica species and arrived on that landmass prior 
to the formation of permanent polar ice sheets \citep{ksepka2006}.  However, our divergence estimates are consistent with global cooling having a profound impact on 
later stages of crown penguin radiation. The Middle Miocene Transition at $\sim$14 Ma marks the onset of a steady decline in 
sea surface temperature, heralding the onset of full-scale ice sheets in Antarctica \citep{zachos2001,hansen2013,knorr2014}. 
Expansion of Antarctic ice sheets may have opened a new environment for \textit{Aptenodytes} and {\it Pygoscelis}, 
the most polar-adapted penguin taxa (including 4 of the 5 species that breed in continental Antarctica). Previous studies have 
placed {\it Aptenodytes} and {\it Pygoscelis} on basal branches of the penguin crown, leading to the hypothesis that crown 
penguins originated in Antarctica and spread to lower latitudes as climate cooled \citep{baker2006}.  However, the geographical 
distribution of stem fossils suggests instead that {\it Aptenodytes} and {\it Pygoscelis} secondarily invaded Antarctica, taking 
advantage of a novel environment \citep{ksepka2006}.  
Our analysis provides additional support for this secondary colonization 
hypothesis by uniting {\it Aptenodytes} and {\it Pygoscelis} as a clade and revealing 
a very recent age for this Antarctic group at 9.8 Ma (Fig.\,\ref{fig:divDates}), indicating they did not radiate until well after permanent ice sheets were established.

\subsection{Morphological clock}

We assume that clock models can be applied to morphological data. A recent study by~\citet{lee2014a} confirms that younger taxa undergo more morphological evolutionary change. Most previous total-evidence or morphological analyses used relaxed clock models for morphology evolution~\citep{Pyron2011,Ronq2012,lee2014a, lee2014b, dembo2015,Zhang2015}. In many studies~\citep{beck2014,lee2014a, dembo2015}, including this study, model comparison analyses favored a relaxed  morphological clock over a strict morphological clock. 

The estimated coefficient of variation for the log-normal distribution of the morphological clock rates in the penguin analysis was 1.15 indicating a high rate variation among the branches. However, in our analysis, the choice of the morphological clock model did not influence much the estimate of the parameter of the primary interest --- the age of the crown radiation.  Using the relaxed clock model as opposed to the strict clock model only slightly shifted the age  toward the past and inflated the 95\% HPD interval (Fig.\,\ref{fig:rootAges}, analyses 7 and 8). 

Many total-evidence analyses inferred implausibly old divergence dates~\citep{Ronq2012,slater2013,Wood2012,beck2014}.  \citet{beck2014} suggested that oversimplified modeling of morphological evolution and a relaxed morphological clock may result in overestimated divergence times. Our analysis did not show this and we, on the contrary, estimated a much younger age of the crown penguin radiation than had been previously estimated. This could be attributed to the large number of stem fossils in our analysis, given that excluding these fossils leads to a much older estimate. The overestimated ages can also be explained by sampling biases \citep{Zhang2015} or using inappropriate tree prior models. Using a birth-death model without sampled ancestors in our analysis slightly shifted the ages of the crown divergences toward the past (Fig.\,\ref{fig:rootAges} and SM, Fig.\,1). 

\subsection{Sampled ancestors}

We examined the total posterior probability of a fossil species to be a sampled ancestor, that is, 
a direct ancestor to other sampled fossil or extant species. If an ancestor-descendant pair is in question, one can also estimate the posterior 
probability of one species to be an ancestor to another species or a group of species as we calculated in the case of the probability of the {\it Spheniscus muizoni} representing an ancestor of extant {\it Spheniscus} radiation. 

The evidence for ancestry comes from morphological data, fossil occurrence times and prior distributions for the parameters 
of the FBD model and morphological evolution model. Here, we used uniform prior distributions except for the net 
diversification rate and morphological evolution rate. 

The analysis of the penguin data set shows a large number of potential sampled ancestors. 
The Bayes factors calculated here showed the ancestry evidence contained in the morphological data. 
The evidence coming from the occurrence times or from all data together remains to be assessed. We hypothesize that the large number of sampled ancestors is due to the temporal pattern of the penguin fossils. 
We additionally analyzed dinosaur~\citep{lee2014b}, trilobite~\citep{congreve2011}, and Lissamphibia~\citep{Pyron2011} datasets with large proportions of missing characters where we only detected up to four sampled ancestors (out of $\sim$ 40--120 fossils; data not shown). An analysis by~\citet{Zhang2015} of Hymenoptera also did not show many sampled ancestors. Thus, the abundance of sampled ancestors in the penguin phylogeny is not likely to be due to the paucity of the morphological data.

\subsection{Further improvements}

Here we used the FBD model, which is an improvement over previously used uniform, Yule, or birth-death models for 
describing the speciation process. However other more sophisticated models may improve the inference or fit better for other data sets. 
The skyline variant of the FBD model \citep{StadKuhn12, gavr2014,Zhang2015}  allows for stepwise changes in rates 
(i.e., diversification, turnover, and fossil sampling) over time. Accounting for the possibility of changes in fossil-sampling rate, 
$\psi$, over time might be important for analyses considering groups of deeply diverged organisms where 
poor fossil preservation may result in underestimates of divergence times if the sampling rate is considered 
constant. Furthermore, models that allow age-dependent \citep{lambert2010,hagen2015} or lineage-dependent 
\citep{Maddison2007,Alfaro2009,alexander2015} speciation and extinction rates
 while appropriately modeling fossil sampling may also improve divergence 
dating and estimation of macroevolutionary parameters.

Another direction of method development is modeling morphological character evolution --- a topic that has sparked numerous 
debates \citep[e.g.,][]{goloboff2003, spencer2013}. 
A recent study by \citet{wright2014} showed that Bayesian methods for estimating tree topologies using morphological data --- 
even under a simple probabilistic Lewis Mk model --- outperform parsimony methods, partly because rate variation is modeled. 
Here, we considered two schemes to assign a number of possible states to a character. The model 
comparison analysis favored the model where the number of possible states is equal to the number of observed 
states in a character. A more accurate modeling would be  to consider each character and assign the number of 
states on the basis of the character description (for example, characters for traits that are either present or absent 
will be assigned two states) or use model averaging within an MCMC analysis where each character is assigned 
different number of states during the MCMC run. 
Moreover, it may also be important to appropriately model ascertainment bias when using the Lewis Mk model.
These extensions include accounting for the absence of invariant and parsimony uninformative characters in the morphological 
data matrix \citep{koch2012}. Importantly, more biologically appropriate models of phenotypic characters are needed to advance 
phylogenetic methods for incorporating fossil data \citep[e.g.,][]{felsenstein2005, revell2014}. The total-evidence method with FBD can also be used to estimate the past 
evolutionary relationship between extinct species where only morphological data is available \citep{lee2014a}.

We assigned age ranges to different fossils in differing ways. Some of the fossil species are known from only one 
fossil specimen and in this case, we assigned the age range based on the uncertainty related to the dating of the layer in 
which the fossil was found. For other species, there are a number of fossils found in different localities. In this case, we 
derived the age interval from probable ages of all specimens. In order to strictly follow the sampling process assumed by the FBD model, it would be necessary  to treat every known fossil specimen individually and include all such specimens into an analysis. Unfortunately, this would lead to enormous data sets with thousands of taxa, most with very high proportions of missing data.
However, in cases where a large number of fossils are known from the same 
locality and are thus very close in age and potentially belong to the same population, this group of fossils 
may be treated as just one sample from the relevant species at that time horizon. Such improved modeling would require devoting considerable effort to
differentiating fossils and recording characters for particular specimens, rather than merging morphological data
from different fossil specimens believed to belong to the same species. This could, however, lead to more accurate inference and 
better understanding of the past diversity.

Finally, our analysis focused on a matrix sampling all fossil penguins represented by reasonably complete specimens. Many poorly known fossil taxa have also been reported, along with thousands of isolated bones. 
Finding a balance between incorporating the maximum number of fossils, which inform sampling rate and time, 
and the computational concerns with adding large numbers of taxa with low proportions of informative characters 
will represent an important challenge for analyses targeting penguins and other groups with extensive fossil records.

We advocate the use of the total-evidence approach with models that allow sampled ancestors when estimating 
divergence times. This approach may offer advantages not only over node calibration methods that rely on 
first analyzing morphological data to identify calibration points and then calibrating phylogenies with {\it ad hoc} prior 
densities, but also over total-evidence methods that do not account for fossils that are sampled ancestors. 
Many recent applications of total-evidence dating have yielded substantially older estimates than node calibration methods \citep[e.g., ][]{Ronq2012,slater2013,Wood2012,beck2014,arcila2014}. Among other explanations~\citep{beck2014, Zhang2015}, using tree priors that do not account for sampled ancestors could have contributed to the ancient dates.

\section{Funding}

AG, DW, and AJD were partially supported by Marsden grant UOA1324 from the Royal Society of NZ. AJD was also supported by a Rutherford Discovery Fellowship from the Royal Society of New Zealand. 
TS is supported in part by the European Research Council under the Seventh Framework Program of the European Commission (PhyPD: grant agreement number 335529). 
TAH and DTK are supported by a US National Science Foundation grant DEB-1556615.

\section{Acknowledgements}

We thank Barbara Harmon and Steph Abramowicz for permission to use the penguin artwork in Figure~\ref{penguinTree}. We also acknowledge the New Zealand eScience Infrastructure (NeSI) for use of their high-performance computing facilities. 
This manuscript was improved after helpful feedback from the editors, Thomas Near and Frank Anderson, and evaluations by Michael Lee, Jeff Thorne, and an anonymous reviewer.

\newpage

\oddsidemargin=0mm
\textwidth=170mm
\topmargin=0pt
\textheight=200mm

\renewcommand{\baselinestretch}{1}

\begin{center}
{\large \textsc{Supplementary Material}}
\end{center}

\begin{center}
\Large{{\bf Bayesian total evidence dating reveals the recent crown radiation of penguins}}
\end{center}

\section{Methods}

\subsection{Prior distributions for parameters}

We used the following prior distributions for the parameters in all analyses except for the sensitivity analyses. 

\vskip2mm

\noindent For the net diversification rate, $d$, we place a log-normal prior distribution with parameters\footnotemark[1]  $M = -3.5$ and $S^2 = 1.5$. The 2.5\% and 97.5\% quantiles of this distribution are $1.6\cdot10^{-3}$ and $0.57$ implying that it well covers the interval (with the lowest 5\% quantile of 0.02 and the largest 95\% quantile of 0.15) estimated in~\cite{Jetz2012} Figure~1.

\footnotetext[1]{Parameters $M$ and $S$ for log-normal distributions here are the mean and standard deviation of the associated normal distributions} 

\vskip2mm

\noindent We place a Uniform(0,1) prior distribution for the turnover, 
$\turnover$, and Uniform(0,1) prior distribution for the fossil sampling proportion, $\fosp$.

\vskip2mm

\noindent For birth, death and sampling rates ($\lambda$, $\mu$, and $\psi$) we used log-normal distributions with parameters\footnotemark[1]  $M = -2.5$ and $S^2 = 1.5$ and  2.5\% and 97.5\% quantiles of $4.34\cdot10^{-3}$ and $1.55$. 

\vskip2mm

\noindent For the time of origin, $t_{or}$, we use the Uniform prior distribution between the oldest sample and 160 Ma because Jetz {\it et al.} estimated the upper bound for the MRCA of all birds to be 150 Ma (\cite{Jetz2012} Figure~1) and Lee {\it et al.}~\cite{lee2014b} estimated the origin of the Avialae clade to about 160 Ma (uncertainty is not reported) in the analysis of morphological data of dinosaurs. 

\vskip2mm

\noindent We used a broad lognormal distribution where the 95\% probability interval spanned $[4\cdot10^{-4}, 0.07]$ which is estimated for extinct dinosaurs in units of changes per character per Myr~\cite{lee2014b} for the rate of morphological evolution, 
$\mu_{morph}$ (the mean of the log-normal distribution of the morphological clock rates). The parameters\footnotemark[1]  of this log-normal distribution are $M = -5.5$ and $S^2 = 2$,  2.5\% and 97.5\% quantiles are $8.11\cdot10^{-5}$ and $0.21$, respectively.   

\vskip2mm 

\noindent $\mu_{mol}$ --- the rate of molecular evolution was assigned a lognormal prior with parameters\footnotemark[1] $M = -3.5$ and $S^2 = 1$ with 5\% and 95\% quantiles of $4.25\cdot10^{-3}$ and $0.21$.   

\vskip2mm 

\noindent We placed a Gamma distribution with parameters $\alpha = 0.5396$ and $\beta = 0.3819$ for the standard deviation of the uncorrelated lognormal relaxed models for both molecular and morphological data.

\vskip2mm 

\noindent The gamma shape parameter on rate-variation across sites was assigned a Uniform[0,10] prior (for both molecular and morphological data). 

\subsection{Data}

The number of characters with different numbers of states is given below:  
\begin{center}
\begin{tabular}{c|c|c|c|c|c|c}
State \# & 2 & 3 & 4 & 5 & 6 & 7\\
\hline
Character \# & 130 &  50 & 13 & 4 & 4 & 1 \\
\end{tabular}
\end{center}
Out of 130 characters with two character states 48 have description `present or absent'.  The number of sites in each region is as follows:
\begin{center}
\begin{tabular}{c|c|c|c|c|c}
Gene & RAG-1 & 12S & 16S & COI & cytochrome b \\
\hline
Site \# & 2682 & 1143 & 1551 & 1105 & 1664 \\
\end{tabular}
\end{center}

\subsection{Fossil ages}

We used the following age ranges for fossil penguins in our analyses (a compact summary is also given in Supplementary Table~\ref{ages}). 

\vskip2mm

\noindent {\bf Taxon:} {\it  Anthropornis grandis.} \\
{\bf Age:} 34--52.5 Ma.  
Specimens are known from Telm 4 and Telm 7 of the La Meseta Formation~\cite{jadwiszczak2006}. Strontium dates from the stratigraphic column of Reguero et al.~\cite{reguero2012} provide ages for the bottom of Telm 4 and the top of Telm 7, which are applied as bounds. 

\vskip2mm

\noindent {\bf Taxon:} {\it  Anthropornis nordenskjoeldi.} \\ 
{\bf Age:} 34--52.5 Ma. The oldest specimens assigned to this species are potentially from Telm 4 or Telm 6 of the La Meseta Formation, while the specimens certainly referable to this species are known from Telm 7~\cite{jadwiszczak2006}. Strontium dates from the stratigraphic column of Reguero et al.~\cite{reguero2012} provide ages for the bottom of Telm 4 and the top of Telm 7, which are applied as bounds. 

\vskip2mm

\noindent {\bf Taxon:} {\it  Archaeospheniscus lopdelli.} \\
{\bf Age:} 26--30Ma. The holotype and only described specimen comes from the Kokoamu Greensand, which is dated to between 26 Ma and 30 Ma (see~\cite{ksepka2012}, Figure~1).   

\vskip2mm

\noindent {\bf Taxon:} {\it  Archaeospheniscus lowei.} \\ 
{\bf Age:} 26--30 Ma.    
All three described specimens come from the Kokoamu Greensand, which is dated to between 26 Ma and 30 Ma (see~\cite{ksepka2012}, Figure~1).  

\vskip2mm

\noindent {\bf Taxon:} Burnside {\it ``Palaeeudyptes''} (this name is a placeholder for a species that has not yet been formally named). \\
{\bf Age:} 36--38.4 Ma.   
This fossil is from a section of the Burnside Formation that can be assigned to the Kaiatan NZ local stage (see~\cite{ksepka2012}, Figure~1).   

\vskip2mm 

\noindent {\bf Taxon:} {\it  Delphinornis arctowskii.} \\
{\bf Age:} 34--41 Ma.   
We considered only tarsometatarsi to be firmly referable to this species. Such specimens are known only from Telm 7 of the La Meseta Formation~\cite{jadwiszczak2006}.  Using the stratigraphic column of Reguero et al.~\cite{reguero2012}, strontium dates for the top and bottom of Telm 7 are applied as bounds. 

\vskip2mm

\noindent {\bf Taxon:} {\it  Delphinornis gracilis.} \\
{\bf Age:} 34--41 Ma.   
We considered only tarsometatarsi to be firmly referable to this species. Such specimens are known only from Telm 7 of the La Meseta Formation~\cite{jadwiszczak2006}.  Using the stratigraphic column of Reguero et al.~\cite{reguero2012}, strontium dates for the top and bottom of Telm 7 are applied as bounds. 

\vskip2mm

\noindent {\bf Taxon:} {\it  Delphinornis larseni.} \\
{\bf Age:} 34--52.5 Ma (species range). \\ 
We considered only tarsometatarsi to be firmly referable to this species. Such specimens are known from Telm 5 and Telm 7 of the La Meseta Formation~\cite{jadwiszczak2006}. Strontium dates for the top of Telm 7 and (because there was no strontium date for the base of Telm 5) the base of Telm 4 (which is very short and certainly $<$ 1 Ma) are used as bounds.  

\vskip2mm

\noindent {\bf Taxon:} {\it  Delphinornis wimani.} \\
{\bf Age:} 34--52.5 Ma.   
We considered only tarsometatarsi to be firmly referable to this species. Such specimens are known from Telm 5 and Telm 7 of the La Meseta Formation~\cite{jadwiszczak2006}. Strontium dates for the top of Telm 7 and (because there was no strontium date for the base of Telm 5) the base of Telm 4 (which is very short and certainly $<$ 1 Ma) are used as bounds.  

\vskip2mm

\noindent {\bf Taxon:} {\it  Duntroonornis parvus.} \\ 
{\bf Age:} 21.7--30 Ma.  
The holotype comes from the Kokoamu Greensand and some referred material comes from the Otekaike Limestone. Ages for the base of the Kokoamu Greensand and top of the Otekaike Limestone are used as bounds (see~\cite{ksepka2012}, Figure~1). 

\vskip2mm

\noindent {\bf Taxon:} {\it  Eretiscus tonnii.} \\
{\bf Age:} 16--21 Ma. 
A possible age range of 16--21 Ma is given based on Figure~2 of Cione et al.~\cite{cione2011}.

\vskip2mm

\noindent {\bf Taxon:} {\it  Icadyptes salasi.} \\ 
{\bf Age:} 35.7--37.2 Ma. This age is based on two radiometrically dated layers that are believed to be correlated with layers above and below the holotype and only reported specimen ~\cite{clarke2007}. 

\vskip2mm

\noindent {\bf Taxon:} {\it  Inkayacu paracasensis.} \\ 
{\bf Age:} 35.7--37.2 Ma.  
This age is based on two radiometrically dated layers that are believed to be correlated with layers above and below the holotype and only reported specimen~\cite{clarke2007}.

\vskip2mm

\noindent {\bf Taxon:} {\it  Kairuku grebneffi.} \\
{\bf Age:} 26--30 Ma.  
The holotype and one referred specimen come from the Kokoamu Greensand, which is dated to between 26 Ma and 30 Ma (see~\cite{ksepka2012}, Figure~1).

\vskip2mm

\noindent {\bf Taxon:} {\it  Kairuku waitaki.} \\
{\bf Age:} 26--30 Ma.  
The holotype and only described specimen comes from the Kokoamu Greensand, which is dated to between 26 Ma and 30 Ma (see~\cite{ksepka2012}, Figure~1).

\vskip2mm

\noindent {\bf Taxon:} {\it  Madrynornis mirandus.} \\ 
{\bf Age:} 9.70--10.3 Ma.  
The holotype and only known specimen is believed to be $10 \pm 0.3$ Ma in age based on strontium dates from the ``Entrerriens'' sequence of the Puerto Madryn Formation where the fossil was collected (see~\cite{hospitaleche2007}).

\vskip2mm

\noindent {\bf Taxon:} {\it  Marambiornis exilis.} \\
{\bf Age:} 34--41 Ma.  We considered only tarsometatarsi to be firmly referable to this species. Such specimens are known only from Telm 7 of the La Meseta Formation~\cite{jadwiszczak2006}.  Using the stratigraphic column of Reguero et al.~\cite{reguero2012}, strontium dates for the top and bottom of Telm 7 are applied as bounds.

\vskip2mm

\noindent {\bf Taxon:} {\it  Marplesornis novaezealandiae.} \\
{\bf Age:} 5.33--15.97 Ma.  This is one of the most difficult penguin fossils to date, because the specimen was found in a loose boulder on a beach that eroded out of a cliff of Great Sandstone deposits. Best estimates for the range of the Greta Siltstone from which the boulder is derived is middle-late Miocene~\cite{feldmann1992}, so a range of 5.33--15.97 is used.

\vskip2mm

\noindent {\bf Taxon:} {\it  Mesetaornis polaris.} \\
{\bf Age:} 34--41 Ma. We considered only tarsometatarsi to be firmly referable to this species. Such specimens are known only from Telm 7 of the La Meseta Formation~\cite{jadwiszczak2006}.  Using the stratigraphic column of Reguero et al.~\cite{reguero2012}, strontium dates for the top and bottom of Telm 7 are applied as bounds.

\vskip2mm

\noindent {\bf Taxon:} {\it  Pachydyptes ponderosus.} \\  
{\bf Age:} 34.5--36 Ma.   All known fossils of this species are all assigned to the Runangan NZ local stage, the bounds of which are used for the age range (see~\cite{ksepka2012}, Figure~1).

\vskip2mm

\noindent {\bf Taxon:} {\it  Palaeeudyptes antarcticus.} \\
{\bf Age:} 30.1--34.5. The holotype and only firmly referred specimen comes from the upper Ototara Limestone, which is assigned, to the Subbotina angiporoides zone of the local Whaingaroan Stage, which is in turn dated to between 30.1--34.5 (see~\cite{ksepka2012}, Figure~1).

\vskip2mm

\noindent {\bf Taxon:} {\it  Palaeeudyptes gunnari.} \\
{\bf Age:} 34--54 Ma.   Specimens are known from Telm 3 and Telm 7 (top) of the La Meseta Formation~\cite{jadwiszczak2006}. Strontium dates for the top of Telm 7 and (because there was no strontium date for the base of Telm 3) the base of Telm 2 (which is very short and certainly $ < 1$ Ma) are used as bounds.

\vskip2mm

\noindent {\bf Taxon:} {\it  Palaeeudyptes klekowski.} \\
{\bf Age:} 34--52.5 Ma (species range).
We included the oldest potentially referable specimen from Telm 5 of the La Meseta Formation as well as the youngest specimens from Telm 7~\cite{jadwiszczak2006} in the age range. Strontium dates for the top of Telm 7 and (because there was no strontium date for the base of Telm 5) the base of Telm 4 (which is very short and certainly $< 1$ Ma) are used as bounds.

\vskip2mm

\noindent {\bf Taxon:} {\it  Palaeospheniscus bergi.} \\
{\bf Age:}  9.7--21 Ma.  Specimens are known from the Gaiman and Chenque Formations, which are contemporaneous, as well as a young record from the Puerto Madryn Formation~\cite{hospitaleche2012}. The age range extends from 21 Ma for the base of Gaiman Formation (Figure~2 of Cione et al. ~\cite{cione2011}) to 9.7 Ma for the Puerto Madryn Formation record~\cite{hospitaleche2007}.

\vskip2mm

\noindent {\bf Taxon:} {\it  Palaeospheniscus biloculata.} \\
{\bf Age:} 16--21 Ma.  A possible age range of 16--21 Ma is given based on Figure~2 of Cione et al. ~\cite{cione2011} representing the duration of the Gaiman Formation.

\vskip2mm

\noindent {\bf Taxon:} {\it  Palaeospheniscus patagonicus.} \\
{\bf Age:} 16--21 Ma.  A possible age range of 16--21 Ma is given based on Figure~2 of Cione et al. ~\cite{cione2011} representing the duration of the Gaiman Formation.
\vskip2mm

\noindent {\bf Taxon:} {\it  Paraptenodytes antarcticus.} \\ 
{\bf Age:} 21--23 Ma.  All known specimens are from the Monte Leon Formation.  Chavez~\cite{chavez2007} showed that reports of this species from the Bahia Inglesia were erroneous. 

\vskip2mm

\noindent {\bf Taxon:} {\it  Perudyptes devriesi.} \\
{\bf Age:}  38--46 Ma.  An age of $\sim$ 42 Ma for the holotype and only known specimen is based on stratigraphic correlations and biostratigraphy~\cite{clarke2007, clarke2010}. Given the lack of directly datable layers at the type locality, we extend this range by 4 Ma in either direction to accommodate uncertainty.

\vskip2mm

\noindent {\bf Taxon:} {\it  Platydyptes novazealandiae.} \\ 
{\bf Age:} 23--26 Ma.   All known specimens come from Oligocene sections of the Otekaike Limestone (see~\cite{ksepka2011}).

\vskip2mm

\noindent {\bf Taxon:} {\it  Platydyptes marplesi.} \\ 
{\bf Age:} 23--30 Ma.  Specimens are known from the Kokoamu Greensand, providing the lower bound and the Otekaike Limestone, providing the upper bound of the age range (see~\cite{ksepka2011}).

\vskip2mm

\noindent {\bf Taxon:} {\it  Pygoscelis grandis.} \\ 
{\bf Age:} 2.5--8.6 Ma.  The oldest specimens are from the Bonebed Member of the Bahia Inglesia Fm. and are from below a layer dated to $7.6 \pm 1.3$ Ma, whereas younger specimens are estimated to be 2.5--4.6 Ma based on biostratigraphy~\cite{walsh2006}.  

\vskip2mm
 
\noindent {\bf Taxon:} {\it  Spheniscus megaramphus.} \\ 
{\bf Age:}  6.3--10 Ma. Stucchi~\cite{stucchi2007} reported this specimens of this species in the Montemar Norte locality and Chavez~\cite{chavez2007} reported material from the Bonebed Member of the Bahia Inglesia Fm. Montemar Norte was estimated to be 10 Ma by Stucchi~\cite{stucchi2008}, providing an upper age limit. Bonebed Member specimens are near a radiometric layer dated to $7.6 \pm 1.3$ Ma, providing the lower bound.

\vskip2mm

\noindent {\bf Taxon:} {\it  Spheniscus muizoni.} \\ 
{\bf Age:} 9--9.2 Ma.  The sole specimen is dated to between 9 and 9.2 Ma based on radiometric dates above and below the specimen~\cite{brand2011}.

\vskip2mm

\noindent {\bf Taxon:} {\it  Spheniscus urbinai.} \\ 
{\bf Age:} 5.7--9.63 Ma. Reported specimens have been recovered from the Sacaco, Sacaco Sur, Montemar, Aguada de Lomas and El Jajuay sites in the Pisco Formation acceding to Stucchi~\cite{stucchi2007}. The 5.7 Ma radiometric date for the top of Sacaco is based on~\cite{ehret2012} and the 9.38 Ma radiometric date for El Jahuay is taken from~\cite{brand2011}.

\vskip2mm

\noindent {\bf Taxon:} {\it  Waimanu manneringi.} \\
{\bf Age:} 60.5--61.6 Ma.  The sole specimen is constrained to between 60.5Ma and 61.6Ma based on biostratigraphy~\cite{slack2006}.

\vskip2mm

\noindent {\bf Taxon:} {\it  Waimanu tuatahi.} \\
{\bf Age:} 56--60.5 Ma.  The most well-constrained fossil of this taxon is dated to 58--60 Ma based on biostratigraphy~\cite{slack2006}. Other referred fossils are less well-constrained and a conservative range would be 56--60.5 is specified with an upper age limit based on the age of {\it Waimanu manneringi} which was collected lower in the stratigraphic column and a lower age limit set to the Paleocene-Eocene boundary, as the fossils are all of Paleocene age.  

\begin{supptable} 
\caption{The stratigraphic age ranges used for phylogenetic analyses of the penguin dataset.}
\begin{center}
\begin{tabular}{c|ll}
\# & Fossil taxon & Age interval \\
\hline
1 & {\it Anthropornis grandis} &  [34, 52.5] \\
2 & {\it Anthropornis nordenskjoeldi} & [34, 52.5] \\
3 & {\it Archaeospheniscus lopdelli} & [26, 30] \\
4 & {\it Archaeospheniscus lowei} & [26, 30] \\
5 & Burnside {\it``Palaeudyptes"} & [36, 38.4] \\
6 & {\it Delphinornis arctowskii} & [34, 41] \\
7 & {\it Delphinornis gracilis} & [34, 41] \\
8 & {\it Delphinornis larseni} & [34, 52.5] \\
9 & {\it Delphinornis wimani} &  [34, 52.5] \\
10 & {\it Duntroonornis parvus} & [21.7, 30] \\
11 & {\it Eretiscus tonnii} &  [16, 21] \\
12 & {\it Icadyptes salasi} & [35.7, 37.2] \\
13 & {\it Inkayacu paracasensis} & [35.7, 37.2] \\
14 & {\it Kairuku grebneffi} & [26, 30] \\
15 & {\it Kairuku waitaki} & [26, 30] \\
16 & {\it Madrynornis mirandus} & [9.7, 10.3] \\
17 & {\it Marambiornis exilis} & [34, 41] \\
18 & {\it Marplesornis novaezealandiae} & [5.33, 15.97] \\
19 & {\it Mesetaornis polaris} & [34, 41]\\
20 & {\it Pachydyptes ponderosus} & [34.5, 36] \\
21 & {\it Palaeeudyptes antarcticus} & [30.1, 34.5] \\
22 & {\it Palaeeudyptes gunnari} & [34, 54] \\
23 & {\it Palaeeudyptes klekowskii} & [34, 52.5] \\
24 & {\it Palaeospheniscus bergi} & [9.7, 21] \\
25 & {\it Palaeospheniscus biloculata} & [16, 21] \\
26 & {\it Palaeospheniscus patagonicus} & [16, 21] \\
27 & {\it Paraptenodytes antarcticus} & [21, 23] \\
28 & {\it Perudyptes devriesi} & [38, 42] \\
29 & {\it Platydyptes novaezealandiae} & [23, 26] \\
30 & {\it Platydyptes marplesi} & [23, 30] \\
31 & {\it Pygoscelis grandis} & [2.5, 8.6] \\
32 & {\it Spheniscus megaramphus} & [6.3, 10] \\
33 & {\it Spheniscus muizoni} & [9, 9.2] \\
34 & {\it Spheniscus urbinai} & [5.6, 9.63] \\
35 & {\it Waimanu manneringi} & [60.5, 61.6] \\
36 & {\it Waimanu tuatahi} & [56, 60.5] \\
\hline
\end{tabular}
\end{center}
\label{ages}
\end{supptable}

\subsection{Validation}

To validate the implementation of the models we simulated 100 trees under the FBD model with parameters: 
$$t_{or} = 60 \quad \text{ and } \quad (d, \turnover, s, \rho) = (0.05, 0.6, 0.2, 0.8).$$ 
We discarded trees with less than five or more than 250 sampled nodes resulting in 96 trees. 
Then we simulated sequences (1000 sites) along the trees for extant samples only under the GTR model with parameters: 
$$ \pi_A = \pi_G = \pi_C = \pi_T = 0.25 \text{ and }$$
$$(\eta_{AG} , \eta_{AC}, \eta_{GT}, \eta_{AT} , \eta_{CG}, \eta_{CT}) = (1, 0.46, 0.45, 0.41, 1.85, 6.22)$$
and an uncorrelated relaxed log-normal model with parameters 
\begin{center}
\begin{tabular}{l}
$\mu_{mol} = 4.36\cdot 10^{-2}$ and \\
$\sigma_{mol} = 0.147$
\end{tabular} 
\end{center}
and morphological characters for all samples (200 characters per sample) assuming each character has ten states under a separate
uncorrelated relaxed log-normal model with parameters 
\begin{center}
\begin{tabular}{l}
$\mu_{morph} = 1.38\cdot 10^{-2}$ and  \\
$\sigma_{morph} = 0.728$.
\end{tabular} 
\end{center}
We chose parameter values that are close to the parameters estimated in the total-evidence analysis of the penguin dataset. 

We analysed simulated morphological data matrix, molecular sequence data and fossil sampling dates to estimate phylogenies, 
FBD model parameters, clock and substitution model parameters. Nine analysis did not converge (one analysis with 239 sampled nodes and eight analyses with too few fossil or extant samples, i.e., fewer than four). The estimates for the remaining 87 converged analysis are summarised in Supplementary Table~\ref{simulations}. 

\begin{supptable}
\caption{The results of the simulation study. The accuracy and precision is very poor for the morphological clock rate. After further removal of four analyses which required more time to converge and that have too few fossil or extant samples (fewer than five) this parameter estimate (numbers in brackets) became more accurate and precise.}
\begin{center}
\begin{tabular}{p{0.1\textwidth} R{0.15\textwidth}R{0.15\textwidth}R{0.15\textwidth}R{0.15\textwidth}R{0.15\textwidth}}
\hline
parameter & true value & mean of relative errors & mean of relative biases & mean of relative 95\% HPD lengths & 95\% HPD accuracy 
\\
\hline
$t_{root}$ & 51 & 0.05 &  $7.4\cdot 10^{-3}$ & 0.22 & 93 \\
$t_{or}$ & 60 & 0.13 & $-4.35\cdot 10^{-3}$ & 0.54 & 95 \\
$d$ & 0.05 & 0.27 & $-0.21$ & 1.57 & 100 \\
$\turnover$ & 0.6 & 0.19 & 0.16 & 1.01 & 98 \\
$s$ & 0.2 & 0.52 & $-0.42$ & $2.4$ & 95 \\
$\rho$ & 0.8 & 0.11 & 0.08 & 0.66 & 100 \\
$\mu_{mol}$ & $4.36\cdot 10^{-2}$ & 0.09 & $-0.06$ & 0.42 & 98 \\
$\sigma_{mol}$ & 0.147 & 0.23 & $-0.03$ & 1.38 & 97 \\
$\mu_{morph}$ & $1.38\cdot 10^{-2}$ & 13.69  & $-13.64$ & 66.39  & 94 \\
& & (0.13) & ($-0.09$)  & (0.62) & (94)  \\
$\sigma_{morph}$ & 0.728 & 0.11 & $-0.04$ & $0.59$ & 99 \\
\hline
\end{tabular}
\end{center}
\label{simulations}
\end{supptable}

\section{Results}

\subsection{Divergence dates}

We performed several different analyses under different model configurations and report the age of the MRCA of modern penguins in Supplementary Table~\ref{tableDivDate}.

\begin{supptable}[!h]
\caption{The ages of the most recent common ancestor of extant penguins for eight analysis of morphological data, 
three analysis of combined data, and one analysis of combined data without stem fossil. The analysis numbers are as in Table~2 of the main text and the abbreviations are explained in the capture to Figure~6 in the main text.}\label{tableDivDate}
\begin{center}
\begin{tabular}{l|c|c}
\hline
Analysis \# & Mean & 95\% HPD \\
\hline 
1 & 13 & [11.43, 14.49] \\
2 & 12.82 & [11.35, 14.39]  \\
3 & 12.45 & [10.7, 14.36] \\
4 & 11.96 & [10.06, 13.83] \\
5 & 12.47 & [10.74, 14.48]\\
6 & 12.5 & [10.75, 14.45] \\
7 & 11.91 & [10.16, 13.76]\\ 
8 & 12.79 & [10.48, 15.7] \\
8+dna R & 12.66 & [9.91, 15.75] \\
8BD+dna R & 13.95 & [10.83, 17.36]\\
8+dna S & 13.46 & [10.21, 16.56] \\  
8+dna R crown only &  22.84 & [14.23, 33.56] \\

\hline
\end{tabular}
\end{center}
\end{supptable}

The estimated divergence dates and HPD intervals from the main analysis (same as Figure~4 in the main text) are given in Supplementary Table~\ref{divDates}. We also performed a total evidence analysis under the birth-death model without sampled ancestors and estimated slightly older divergence dates (Supplementary Figure~\ref{BDvsSABD}).  

\begin{supptable}[!h]
\caption{Estimated divergence dates for extant penguin lineages inferred from the total-evidence analysis including all fossil and extant taxa. Node numbers correspond to the numbers in circles in Figure~4 of the main text.}
\begin{center}
\begin{tabular}{c c c}
\hline
Node & Mean & 95\% HPD \\
\hline 
1 & 12.66 & [9.91, 15.75] \\
2 & 11.19 & [7.92, 12.97] \\
3 & 10.34 & [7.14, 11.98] \\
4 & 9.83 & [5.84, 14.04] \\
5 & 6.12 & [3.3, 9.04] \\
6 & 5.08 & [3.27, 6.86] \\
7 & 3.47 & [1.68, 5.27] \\
8 & 2.42 & [1.49, 3.33] \\
9 & 2.2 & [1.11, 2.85] \\
10 & 2.1 & [1.05, 2.6] \\
11 & 1.65 & [0.99, 2.32] \\
12 & 1.63 & [0.74, 2.49] \\
13 & 1.06 & [0.63, 1.88] \\
14 & 1.06 & [0.54, 1.6] \\
15 & 0.77 & [0.35, 1.23] \\
16 & 0.76 & [0.38, 1.16] \\
17 & 0.44 & [0.19, 0.71] \\
18 & 0.32 & [0.09, 0.58] \\
\hline
\end{tabular}
\end{center}
\label{divDates}
\end{supptable}

\begin{suppfigure}
\caption{Maximum clade credibility trees with common ancestor ages obtained after removing fossil lineages from the posterior trees (see section ``Summarising trees" in the main text) and 95\% HPD intervals for divergence times from total-evidence analyses under the birth-death model with sampled ancestors (blue phylogeny) and without (red phylogeny). The divergence ages estimated under the birth-death model without sampled ancestors are slightly older.}
\begin{center}
\includegraphics[width=\textwidth]{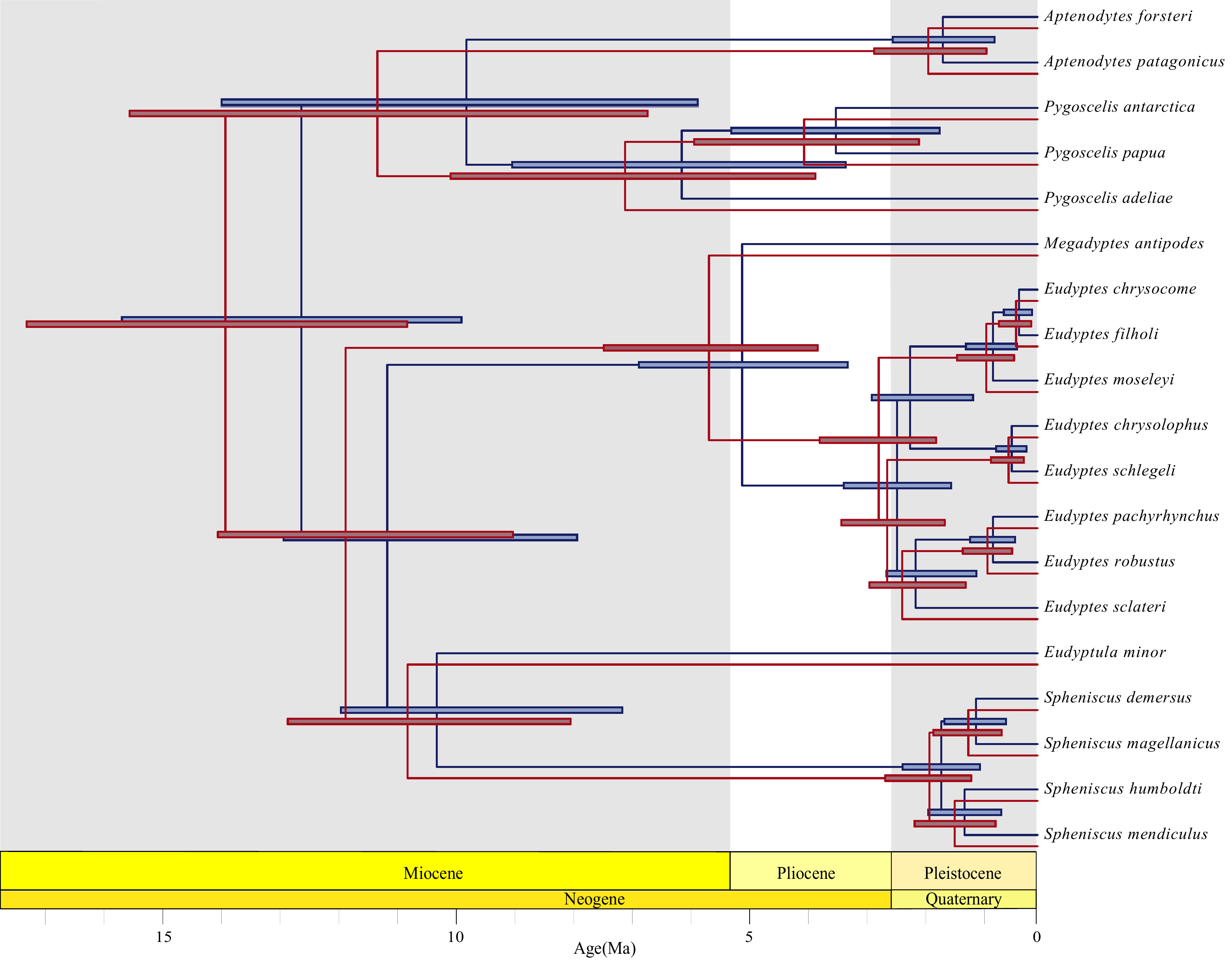}
\end{center}
\label{BDvsSABD}
\end{suppfigure}

\subsection{Sensitivity to the choice of prior distributions for the net diversification rate}

Given enough comparative data of fossil and living samples, the FBD model allows us to estimate all hyper-parameters~\cite{gavr2014}.
Thus, the true value of a parameter will be in the 95\% HPD interval 95\% of the time for analyses with uninformative priors for the FBD model hyper-parameters. Broad uninformative priors (for diversification rate which takes values from zero to infinity, for example) increase the time required for the convergence of an MCMC run. This is not an issue for the penguin dataset  as it is relatively small. However, since we estimated marginal likelihoods for different models, the convergence time of a single analysis became important. To reduce this time we used a log-normal prior distribution for the net diversification rate. We investigated the impact of different prior distributions on the estimate of this parameter and on the parameter of the primary interest --- the age of the MRCA of extant penguins. The results are shown in Supplementary Figure~\ref{pr_post} and Supplementary Figure~\ref{root_ages_dif_p}.

\begin{suppfigure}
\caption{Prior and posterior probability densities for the net diversification rate for analysis~1 from Table~2 in the main text
under four different prior distributions.}
\begin{center}
\includegraphics[width=0.7\textwidth]{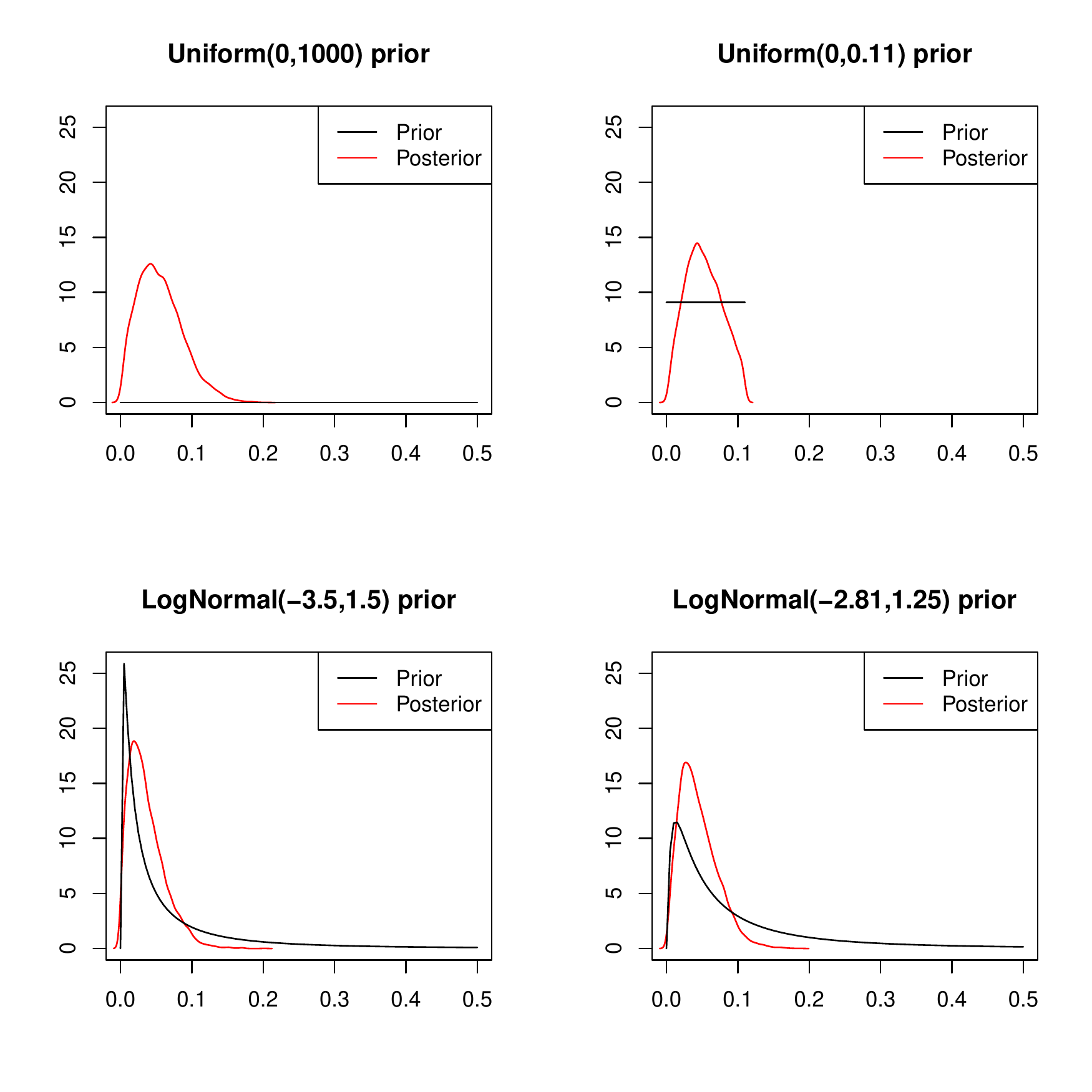}
\end{center}
\label{pr_post}
\end{suppfigure}

\begin{suppfigure}
\caption{The estimates of the MRCA age of extant penguins for analysis~1 from Table~2 in the main text
under four different prior distributions for the net diversification rate. The $x$-axis is labeled with the prior distributions used for the net diversification rate. These different prior distributions have little effect on the age of the MRCA.}
\begin{center}
\includegraphics[width=0.5\textwidth]{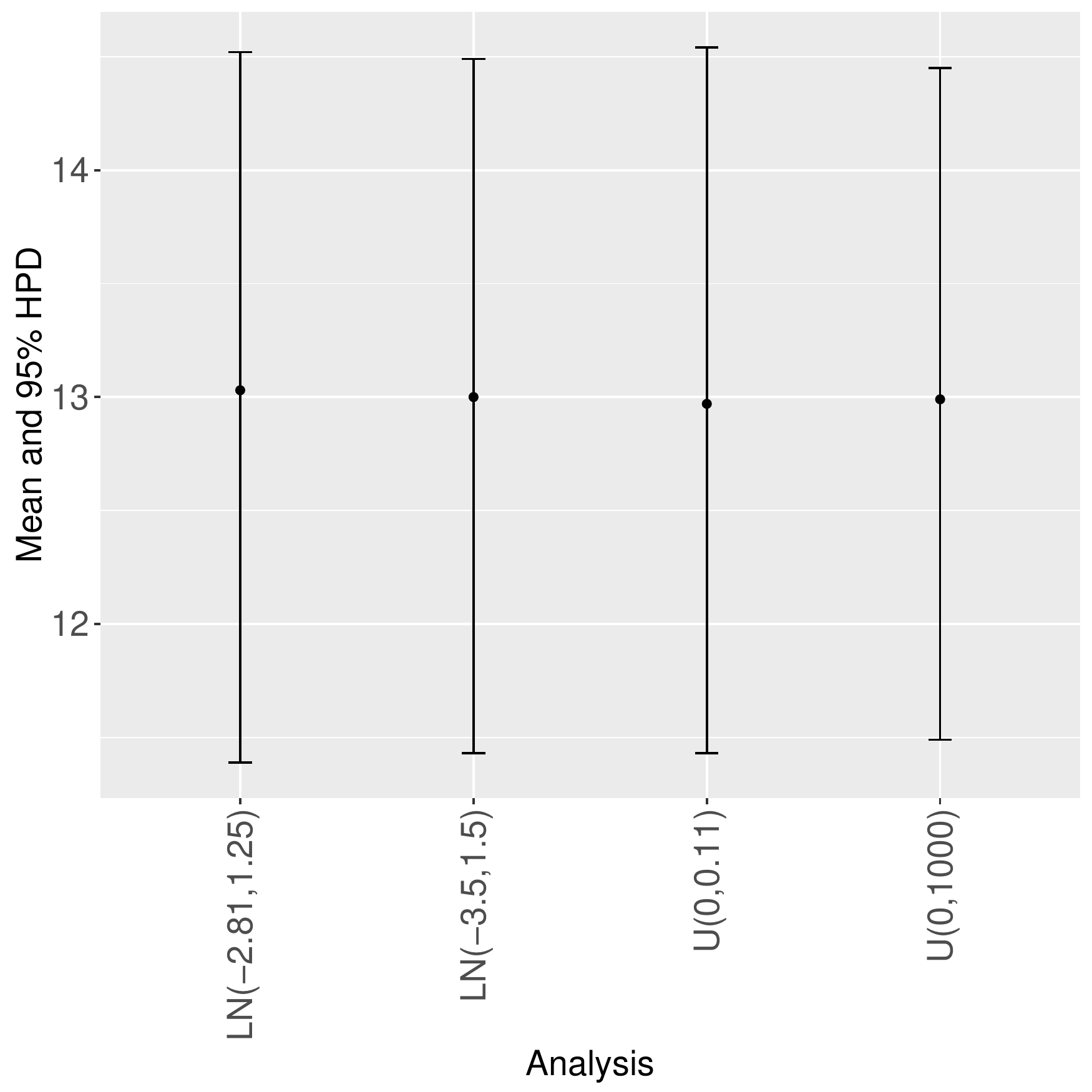}
\end{center}
\label{root_ages_dif_p}
\end{suppfigure}

\bibliographystyle{sysbio}

\bibliography{Total-evidenceWithSA}

\end{document}